\documentclass[fleqn,usenatbib]{mnras}

\usepackage{newtxtext,newtxmath,color}
\usepackage{threeparttable}
\usepackage{multirow}
\usepackage{booktabs}
\usepackage{ulem}
\usepackage{makecell}

\usepackage[T1]{fontenc}

\DeclareRobustCommand{\VAN}[3]{#2}
\let\VANthebibliography\thebibliography
\def\thebibliography{\DeclareRobustCommand{\VAN}[3]{##3}\VANthebibliography}

\usepackage{graphicx}	
\usepackage{subcaption}
\usepackage{amsmath}	
\usepackage{amssymb}	
\usepackage{cellspace}
\usepackage{makecell}
\usepackage{longtable}
\title[Influences of SP sources on EoR detections]{Influence of sources with a spectral peak in the detection of Cosmic Dawn and Epoch of Reionization}

\author[M. He et al.]{
Mengfan He$^{1,2,3}$\thanks{E-mail: mfhe@bao.ac.cn},
Qian Zheng$^{2,4}$\thanks{E-mail: qzheng@shao.ac.cn},
Quan Guo$^{2,4}$\thanks{E-mail:guoquan@shao.ac.cn},
Huanyuan Shan$^{2,3,4}$,
Zhenghao Zhu$^{2}$,
Yushan Xie$^{2}$,
\newauthor{Yan Huang$^{1}$ and Feiyu Zhao$^{2,3}$}
\\
$^{1}$National Astronomical Observatories, Chinese Academy of Sciences,Beijing 100012, China\\
$^{2}$Shanghai Astronomical Observatory, Chinese Academy of Sciences, 80 Nandan Road, Shanghai 200030, China\\
$^{3}$School of Astronomy and Space Science, University of Chinese Academy of Sciences, Beijing 100049, P.R. China\\
$^{4}$Key Laboratory of Radio Astronomy and Technology , Chinese Academy of Sciences, A20 Datun Road, Chaoyang District, Beijing, 100101, P. R. China}

\date{Accepted XXX. Received YYY; in original form ZZZ}

\pubyear{2015}

\begin{document}
\label{firstpage}
\pagerange{\pageref{firstpage}--\pageref{lastpage}}
\maketitle


\begin{abstract}
Foreground removal is one of the biggest challenges in the detection of the Cosmic Dawn (CD) and Epoch of Reionization (EoR). Various foreground subtraction techniques have been developed based on the spectral smoothness of foregrounds. However, the sources with a spectral peak (SP) at Megahertz may break down the spectral smoothness at low frequencies ($<$ 1000\,MHz). 
In this paper, we cross-match the GaLactic and Extragalactic All-sky Murchison Widefield Array (GLEAM) extragalactic source catalogue with three other radio source catalogues, covering the frequency range 72\,MHz--1.4\,GHz, to search for sources with spectral turnover. 
4,423 sources from the GLEAM catalogue are identified as SP sources, representing $\approx$ 3.2 per cent of the GLEAM radio source population.
We utilize the properties of SP source candidates obtained from real observations to establish simulations and test the impact of SP sources on the extraction of CD/EoR signals.
We statistically compare the differences introduced by SP sources in the residuals after removing the foregrounds with three methods, which are polynomial fitting, Principal Component Analysis (PCA), and fast independent component analysis (FastICA). 
Our results indicate that the presence of SP sources in the foregrounds has a negligible influence on extracting the CD/EoR signal. 
After foreground subtraction, the contribution from SP sources to the total power in the two-dimensional (2D) power spectrum within the EoR window is approximately 3 to 4 orders of magnitude lower than the CD/EoR signal. 

\end{abstract}


\begin{keywords}
general — cosmology: dark ages, reionization, first stars – instrumentation: interferometers – methods: observational
\end{keywords}



\section{Introduction}
The redshifted 21 cm signal of neutral hydrogen is a powerful tool to probe the CD/EoR signal, which is supposed to be detected in the low frequency range from 50 to 200\,MHz. The CD/EoR signal is extremely faint and deeply buried under the bright foregrounds dominated by the emission of extragalactic radio sources and our Galaxy. 
In order to extract such faint CD/EoR signals, a number of parametric algorithms are developed to remove or avoid the foreground components. Many of these algorithms rely on the assumption of spectrum smoothness \citep[e.g.][]{Santos2005, Wang2006, McQuinn2006, Bowman2006, Jelic2008, Gleser2008, Liu2009,Petrovic_Oh2011,Harker2010, Sims2016, Chapman2016}. Even for nonparametric methods, with which subtracting foregrounds is performed with minimal prior assumptions on the frequency dependence of foreground components, such as Principal Component Analysis \citep[PCA;][]{Cunnington2021}, fast independent component analysis \citep[FastICA;][]{Alonso2015}, Generalised Morphological Component Analysis \citep[GMCA;][]{Carucci2006}, the smoothness of foreground frequency spectra remains a crucial factor for effectively separating the foreground and CD/EoR signal. However, the presence of sources with  a spectral peak (SP) at megahertz range, may disrupt the smoothness of the foreground.

The Gigahertz-Peaked Spectrum (GPS), High-Frequency Peaked (HFP), and Compact Steep Spectrum (CSS) are three classes of foreground sources with turnovers in their spectra \citep{O'Dea2021}. Among them, the spectra of GPS and HFP sources always turn over at a few gigahertz. CSS sources are a class of radio sources with properties similar to GPS and HFP sources, but their peak frequencies are lower. 
Sources with a Megahertz spectral peak are thought to be a combination of nearby CSS sources and GPS sources at high redshifts, resulting in turnover frequencies that have shifted below a gigahertz due to cosmological evolution \citep{Coppejans2015}. The spectral turnovers of GPS and CSS sources can be attributed to self-absorption mechanisms such as synchrotron self-absorption (SSA) and free-free absorption (FFA).

Multi-frequency observations are required for identifying the SP sources, especially at very low frequencies from a few tens to hundreds MHz. There are many radio interferometers built in radio quiet areas, such as 21 CentiMeter Array \citep[21CMA;][]{Zheng2016ApJ}, Murchison Widefield Array \citep[MWA;][]{Cathryn2020}, the Giant Metrewave Radio Telescope \citep[GMRT;][]{Paciga2013} and the LOw-Frequency ARray \citep[LOFAR;][]{Patil2017, Mertens2020} etc., providing us with valuable information about these sources at low-frequency radio band. 

Previous studies on peaked-spectrum sources focused mainly on searching for or investigating their properties and physical mechanisms. \citet{Mhaskey2019} extracted a list of fifteen extremely inverted spectrum extragalactic radio sources from the northern sky Westerbork Northern Sky Survey (WENSS) and TIFR GMRT Sky Survey (TGSS-ADR1) radio surveys. In the southern sky, \citet[][]{Callingham2017} provides a sample of 1,483 extragalactic peaked-spectrum radio sources based on the GLEAM survey (referred to as the Callingham's sample hereafter), including 261 GPS sources with spectral peaks above 843\,MHz/1.4\,GHz and 1,222 sources with peaks between 72\,MHz and 1.4\,GHz.
In addition to the Callingham's sample, they also provide another sample that contains 116 sources displaying a convex spectrum (referred to as the Convex-Sample hereafter), and a separate sample consisting of 36 sources that peak below 72\,MHz.
Their study did not find a correlation between the peak frequency and the redshift.
\citet{Keim2019} studied six sources with Megahertz-Peaked Spectrum from the GLEAM survey and found that spectral peaks of a fraction of sources could be well described by the FFA model. 

\citet{Zheng2012} investigated the impact of SSA sources on the detection of CD/EoR signals. They developed a phenomenological model to characterise the spectra of SSA sources. 
By simulating extragalactic radio sources with and without the inclusion of SSA sources over a sky area of $10^{\circ} \times 10^{\circ}$, they concluded that the influence of SSA on the detection of CD / EoR signals is likely negligible. 
With the development of low-frequency radio telescopes, more reliable tests based on observation data can be performed. 
This will provide a more accurate assessment of the impact of spectral turnover sources on the detection of CD / EoR signals and will further validate the findings of \citet{Zheng2012}.

In this work, we perform a multifrequency study of radio sources and identify the SP source candidates. We further build up simulations based on observed SP sources and estimate their influence on the measurement of CD/EoR signals (hereafter referred to as the EoR signals), including both power spectrum and imaging. 
This paper is organised as follows. In Section~\ref{sec:data}, we introduce the source catalogues used in this work. Section~\ref{sect:Candidates} presents the method and the results of SP sources selection. In Section~\ref{sect:results}, we show the results of the test of the influence of the SP sources on the removal of the background CD / EoR. The discussion and conclusion are given in Section~\ref{sec:discussion}.

\section{Data}
\label{sec:data}
In order to detect spectral turnovers of radio objects (hereafter referred to as SP sources), observations with wide frequency coverage are required. Therefore, we take advantage of six catalogues covering 72\,MHz to 1.4\,GHz to search for candidates with peaks and study their spectral properties.


GLEAM survey \citep[][]{Wayth2015} is carried out by MWA and covers an area of 5,113\,$\rm deg^2$ including almost the entire sky south of declination +25 degree \citep[][]{Wayth2015}. MWA is a low-frequency radio interferometer located in Western Australia. 
It consists of 4,096 bowtie dipole antennae in 256 tiles, spread over several kilometres, resulting in an angular resolution of approximately 2 arcminutes (at 216\,MHz).
In this work, we use the GLEAM extragalactic source catalogue\footnote{\url{https://cdsarc.cds.unistra.fr/viz-bin/Cat?VIII/100}}, comprising 307,455 radio sources \citep[][]{Hurley2017} with twenty contemporaneous flux density measurements between 72 and 231\,MHz (hereafter referred to as GLEAM band), making it a valuable resource for investigating the spectral characteristics of extragalactic sources at low frequencies.
The completeness of GLEAM is almost 100 per cent at 1\,Jy (for the declination $\delta<18.5^{\circ}$). The flux density error in each subband is taken as the sum in quadrature of the Gaussian fitting error (calculated by \textsc{Aegean}\footnote{\url{https://github.com/PaulHancock/Aegean}}).

For the first time, we use the Rapid ASKAP Continuum Survey \citep[RACS;][]{ASKAP2021} to search for candidates of the peaked spectrum. 
RACS is the first large sky survey using the Australian Square Kilometre Array Pathfinder \citep[ASKAP;][]{ASKAP2021}, a radio synthesis array also located in western Australia, comprising 36 12m dish antennas spread over 6\,kilometres. 
The survey covers the sky south of the +41$^{\circ}$ declination. 
It is the deepest Southern sky radio survey at the corresponding frequency range so far, using 903 individual pointings with 15-minute observations. 
We use the first release of the RACS Stokes I catalogue, which contains 2,123,638 radio sources at a central frequency of 887.5\,MHz with a common resolution of 25\,arcsec, and covers a large contiguous region in the declination range from -$80^{\circ}$ to +$30^{\circ}$. 
It is a comparatively complete Southern sky catalogue and can be well matched to the GLEAM catalogue. We take the "E\_Total\_flux\_Source" error as the flux-density error for the RACS data, which is a combination of the error on the total flux density derived by summing in quadrature the error from PyBDSF \citep[][]{Mohan&Rafferty2015} with the errors of flux density from Eq.~7 of \citet[][]{McConnell2020}.

\begin{table*} 
\centering
\caption{Survey properties of GLEAM, RACS, VLSS, SUMSS and NVSS. The `Position accuracy' column shows the position accuracy of the catalogues. `Offset' refers to the positional offset between RACS measurements and other catalogues, such as SUMSS and NVSS \citep{ASKAP2021}. }
\label{Table:catalogues}
\begin{tabular}{c|c|c|c|c|c|c}
    \hline
         Catalogue & Telescope & Frequency (MHz) & Survey region & Resolution & Position accuracy & Reference \\ 
         \hline
         VLSSr      & VLA  &   74     & $\delta > -30^{\circ}$ & $75^{\prime \prime}$  & $30^{\prime \prime}$&\citet{VLSSr2014}\\
         MRC        & Molonglo &   408  &  $-85^{\circ}< \delta \tnote{*} <+18.5^{\circ}$ &  &  &\citet[][]{Large1981, Large1991}\\
        GLEAM     & MWA  &  72-231  & $\delta < +30^{\circ}$   & $\approx 2^{\prime}$  & $\approx30^{\prime \prime}$& \citet[][]{Wayth2015}\\
         SUMSS     & MOST &   843    & $\delta \le -30^{\circ}$  & $\approx 45^{\prime \prime}\times 45^{\prime \prime}$ cosec|$\delta$| & better than $10^{\prime \prime}$&\citet[][]{SUMSS2008}\\
         RACS      & ASKAP&   887.5  & $-80^{\circ}< \delta <+30^{\circ}$ & $25^{\prime \prime}$  & |Offset| $\le 0.8^{\prime \prime}$&\citet[][]{ASKAP2021}\\
         NVSS      & VLA  &   1400   & $\delta > -40^{\circ}$  &  $\approx 45^{\prime \prime}$ & within 1$^{\prime \prime}$&\citet[][]{NVSS1998} \\
         \hline
\end{tabular}
\begin{tablenotes}
\footnotesize
\item[*] The MRC survey region is expressed in J1950 coordinates.
\end{tablenotes}
\end{table*}

In addition to these two catalogues, we also include data from the NRAO VLA Sky Survey \citep[NVSS;][]{NVSS1998} and the Sydney University Molonglo Sky Survey \citep[SUMSS;][]{SUMSS2008} to help us identify the SP sources. Once an SP source is identified, it is cross-matched with the Very Large Array Low-frequency Sky Survey Redux catalogue \citep[VLSSr;][]{VLSSr2014} and the Molonglo Reference Catalogue \citep[MRC;][]{Large1981, Large1991} to help us determine the accuracy of the spectral fit to the GLEAM and NVSS/RACS/SUMSS data. 
The details of these catalogues are summarised in Table~\ref{Table:catalogues}. 
Note that, for flux density error, besides the fitting error (rms) given by the catalogues, additional calibration errors should be considered when combining measurements from different observations. 
In our work, we add the internal systematic uncertainties to the flux density measurements from the GLEAM catalogue, i.e., 2 per cent for sources distributed in $-72^{\circ}\le{\rm \delta}< 18.5^{\circ}$ and 3 per cent for other sources \citep[][]{Wayth2015}.
In addition to the GLEAM data, we add an additional systematic error 4 per cent for SUMSS measurements, while 4 per cent for NVSS \citep[][]{Callingham2017} and 4 per cent for RACS \citep[][]{ASKAP2021} to account for calibration errors between different catalogues. 
The final uncertainty for each source is the combination of fitting errors ($\sigma_{\rm rms}$) given by the catalogues and the systematic uncertainties ($\sigma_{\rm sys}$) we added, that is, $\sigma = \sqrt{{\sigma_{\rm rms}}^2+\sigma_{\rm sys}^2}$.

\section{Source Candidates}
\label{sect:Candidates}

\subsection{Catalogues Cross-Match}

In order to collect multi-waveband data for each radio source from the different catalogues, cross-matching between catalogues is required.
Considering that the GLEAM catalogue is the most comprehensive in the southern sky for frequencies below 300 MHz and includes multiflux density measurements from 72 to 231 MHz, we use the GLEAM extragalactic catalogue as the primary reference, and individually cross-matched it to SUMSS, RACS, and NVSS using an angular separation cutoff of 2$^{\prime}$20$^{{\prime}{\prime}}$. 
Cross-matching is performed using the Positional Update and Matching Algorithm \footnote{\url{https://github.com/JLBLine/PUMA}} \citep[PUMA;][]{Line2017}. 
Following \citet[][]{Callingham2017}, we chose sources classified as \texttt{isolated} by PUMA, which means that the source chosen in each catalogue does not have neighbours within an angular radius of 2$^{\prime}$20$^{{\prime}{\prime}}$. For these sources to be further accepted, all matched sources must be located within 1$^{\prime}$10$^{{\prime}{\prime}}$ of the source position from the GLEAM catalogue or have a probability of cross-matching the position greater than 0.99. 
In cases where multiple counterparts were found in a single catalogue, all will be rejected to avoid potential mismatches caused by unresolved sources. The matched results are summarised in Table~\ref{GLEAM_matching}, in which the last column shows the number of sources that have only one counterpart. 
300,421 out of 2,123,638 sources in the RACS catalogue are found to match about 97.7 per cent of the GLEAM objects. If we cross-match two catalogues only using the criterion of positions disparities within 2$^{\prime}$20$^{{\prime}{\prime}}$, the matching rate can reach 99.4 per cent for sources with declation in the range $-80^{\circ}\le \delta \le+30^{\circ}$. 
The additional 0.3 per cent of unmatched sources can be attributed to slight differences in the sky coverage of RACS and GLEAM at the edges of the surveys.
The remaining 0.3 per cent can be attributed to differences in angular resolution, noise, and sensitivity between the two surveys, resulting in greater positional discrepancies between the two catalogues or the fact that some sources were not detected by RACS. 

\begin{table}
\centering
    \begin{tabular}{c|c|c|c|c}
    \hline
         catalogue &  Frequency(MHz) & $N_{\rm cat}$ & $N_{\rm match}$ & $N_{\rm 1-match}$\\ 
         \hline
         VLSS      &   74     & 68,311    & 44,311  & 44,028\\
         GLEAM     &  72-231  & 307,455   &        &   \\
         MRC       &   408    & 12141     & 10,039  & 10,038\\
         SUMSS     &   843    & 211,050   & 100,233 & 94,583\\
         RACS      &   887.5  & 2,123,638 & 300,421 & 275,588\\
         NVSS      &   1400   & 1,773,484 & 233,861 & 206,774\\
\hline
    \end{tabular}
    \caption{Catalogue matching results. The column $N_{\rm cat}$ is the number of sources in each catalogue, and column $N_{\rm match}$ shows the number of sources matched to the GLEAM catalogue. The $N_{\rm 1-match}$ column shows the number of sources that have only one matched counterpart in the corresponding catalogue.}
    \label{GLEAM_matching}
\end{table}

\subsection{Models}
\label{sect:SA}
In this work, four spectral models are used to fit the spectra of sources. SP sources candidate selection is based on fitting results of these models.

Ordinary radio source spectra can be well-fitted by the standard non-thermal power-law model (PL model), which is described by the following equation:
\begin{equation}
S_v=av^{\alpha},
\label{eq:PL}
\end{equation}
where, $S_\nu$ represents the flux density, $a$ describes the amplitude of the spectrum, and $\alpha$ represents the spectral index.

We use the generic curved (GC) model and the homogeneous FFA model to describe peaked spectra. These two models are collectively dubbed SP models in the following context. The GC model can be written as \cite[][]{Callingham2017}:
\begin{equation}
    S_v= \frac{S_P}{(1-e^{-1})}(1-e^{-(v/v_{p})^{\alpha_{\rm thin}-\alpha_{\rm thick}}})(\frac{v}{v_{\rm p}})^{\alpha_{\rm thick}},
\label{eq:GC}
\end{equation} 
where $S_P$ is the flux density at frequency $v_{\rm p}$ (note that $v_{\rm p}$ is related to peak frequency, but not exactly the peak frequency), $\alpha_{\rm thin}$ and $\alpha_{\rm thick}$ are the spectral indices of optically thin and optically thick regimes of the spectrum, respectively. It can be reduced to a homogeneous SSA source when $\alpha_{\rm thick}=2.5$. Note that this model is built to describe the curved spectrum of a source, but cannot be used to assess whether SSA or FFA is responsible for such spectrum \citep[][]{Callingham2017}. Power-law slopes are required below and above the peak in the spectrum in Eq.~\ref{eq:GC}. However, for some sources, their spectra turn out to be more complex and cannot be well fitted by Eq.~\ref{eq:GC}. Therefore, \citet[][]{Callingham2017} gave a homogeneous FFA model for those sources, defined as:
\begin{equation}
    S_v= av^{\alpha}e^{(v/v_{\rm p})^{-2.1}},
\label{eq:FFA}
\end{equation}  
where $a$, $\alpha$ and $v_{\rm p}$ are free parameters. This equation can be used to model the spectra, which are exponentially decreasing below the peaks. More details about the GC model and the FFA model are given in \citet[][]{Callingham2017}. 

In addition to the power-law (PL) and SP models, we also employ the curved power-law model (CPL model) \citep{Callingham2017} to quantify spectral curvature. The CPL model is defined as:
\begin{equation}
S_v=av^\alpha e^{q\left({\rm ln}v\right)^2}.
\label{eq:CPL}
\end{equation}
Here, $q$ characterizes the spectral curvature. When $|q|\geq 0.2$, the spectrum exhibits a significant curvature \citep{Duffy2012}.

When performing fitting, it is crucial to account for the known correlations between sub-band flux densities within the GLEAM band \citep[see Section 5.4 of][]{Hurley2017}. These correlations between the flux densities of the subband are modelled using the Matérn covariance function \citep[][]{Rasmussen2006}. 
We use a \textsc{Python} package \textsc{George}\footnote{\textsc{George} 0.3.0: \url{http://dfm.io/george/current/}} to construct the Matérn covariance. The fitting is performed using \textsc{SciPy} \footnote{\textsc{SciPy} 1.6.3: \url{https://docs.scipy.org/doc/scipy-1.6.3/reference/}}, taking into account the Matérn covariance constructed by \textsc{George}.
We calculate the p-value of the F-test from the fitting results of the SP model and the power-law (PL) model, as follows,
\begin{equation}
\hat{F} = \frac{{(\chi^2_1 - \chi^2_2) / ({\rm dof_1} - {\rm dof_2})}}{{\chi_2 / {\rm dof_2}}} 
\end{equation}

\begin{equation}
Prob(F\ge \hat{F}) = \text{{CDF}}(F({\rm dof_1} - {\rm dof_2}, {\rm dof_2}))
\end{equation}
then p-value =  1 - $Prob$, here $\chi$ is calculated from the difference between best-fitting and observational data, the dof refers to the degree of freedom, while the subscripts 1 and 2 represent Model 1 and Model 2, respectively.
A smaller p-value indicates that the SP model provides a better fit to the data compared to the PL model. 
For each source, the best-fitting SP model is the one among the GC and FFA models with the smaller normalized $\chi^2$, where the reduced $\chi^2$ is defined as $\chi^2/\rm{dof}$, and $\rm{dof}$ represents the degrees of freedom.
The frequency at which the maximum flux density occurs in the SP model is denoted as $V_{\rm p}$, and the flux density at $V_{\rm p}$ is denoted as $S_{\rm peak}$. 
Furthermore, we fit the power-law (PL) model to only the flux density measurements from GLEAM. The spectral index obtained from this fitting is denoted as $\alpha_{_{\rm low}}$, which helps us in selecting the peaked spectral sources in the next section.
Unless otherwise specified, the values of $\alpha$, $V_{\rm p}$, $v_{\rm p}$, $S_{\rm peak}$, and the p-value are calculated from the fitting results of the SP model and the PL model, which are obtained by fitting the matched spectrum. 
The values of $q$, $\Delta_q$, and $\alpha_{_{\rm low}}$ are derived solely from the fitting of the GLEAM data points.


\begin{table*}
    \centering
    \begin{tabular}{cccc}
    \hline
         Selection step & Selection criterion & $\rm N_{sources}$& $\rm N_{unresolved}$\\
         \hline
         \multirow{2}*{S1} & $\delta \ge -80^{\circ}$  & 304,942 & \\
         & with 8 or more flux density measurements (combined) with a SNR $\ge$3 & 246,202\\
         \hline 
         S2a & $\alpha \ge 0.1$, have matches from SUMSS, RACS, or NVSS, $\alpha_{_{\rm low}}\ge0.1\, {\rm or}\, (q\ge0.2$ and $\Delta_q\le0.2)$ & 810 & 734\\
         S2b & $\alpha < 0.1$, $V_{\rm p} \ge 130 $MHz, p-value$\le$ 0.005 & 2,745 & 2,428 \\
         S2c & $72 \,{\rm MHz}\le V_{\rm p}<130\,{\rm MHz}$, $q\le-0.2$,  $\Delta_q\le0.2$, p-value $\le$ 0.005 & 792& 765\\
         S2d &  $q\le-0.5$, $(q+\Delta_q)\le-0.1$, $v_{\rm p}< 600$\,MHz,  p-value$\le$ 0.005 & 1,174& 1,063\\
         \hline
         Total & SP sources selection & 4,423& 3,927\\
    \hline
    \end{tabular}
    \caption{A summary of SP sources candidates selection criterion and number of sources after each selection step. The $\rm N_{sources}$ column shows the number of sources, while the $\rm N_{unresolved}$ shows the number of resolved sources. Note that these four samples are not mutually exclusive.}
    \label{tab:select_routine}
\end{table*}

\subsection{SP Source Selection}
\label{subsec:SA_selection}

\begin{figure}
\centering
\includegraphics[width=0.98\hsize]{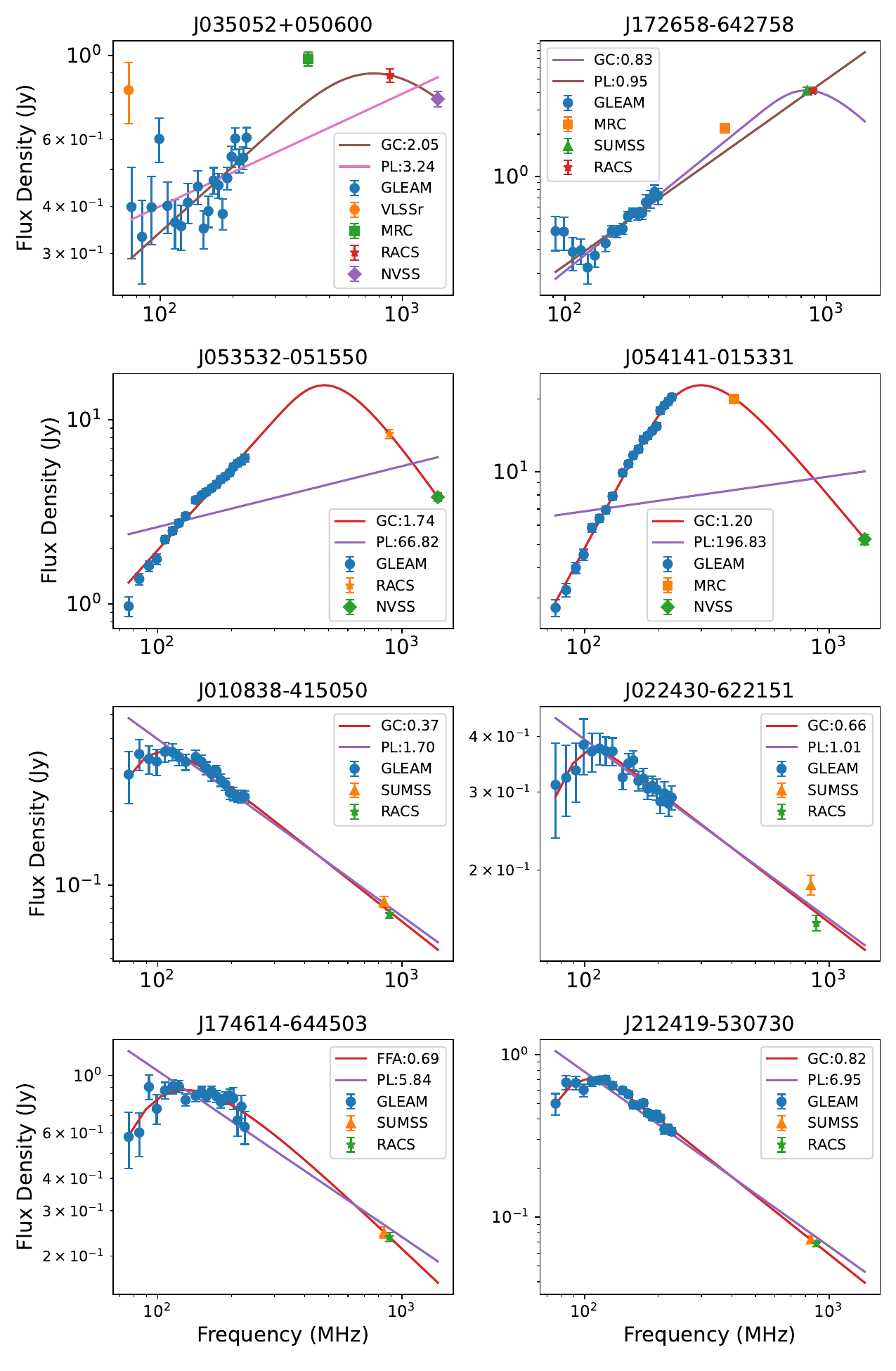}
\caption{Example spectra for sources identified via our selection process. These sources were not identified by \citet[][]{Callingham2017}. The figures, arranged from top to bottom, display the spectra of sources from the S2a, S2b, S2c, and S2d samples, respectively. The sources shown in the top two rows are resolved sources. The blue data points with error bars represent the flux density measured by GLEAM. Symbols corresponding to data points from VLSSr, MRC, SUMSS, and NVSS can be found in the legend of each figure. The lines in the figures represent spectra obtained by fitting the data with the SP model and the PL model (as indicated in the labels for each figure). The number following each legend with 'GC,' 'FFA,' or 'PL' is the reduced $\chi^2$ value for the respective model.}
\label{candis}
\end{figure}

\begin{figure}
\centering
\includegraphics[width=0.95\hsize]{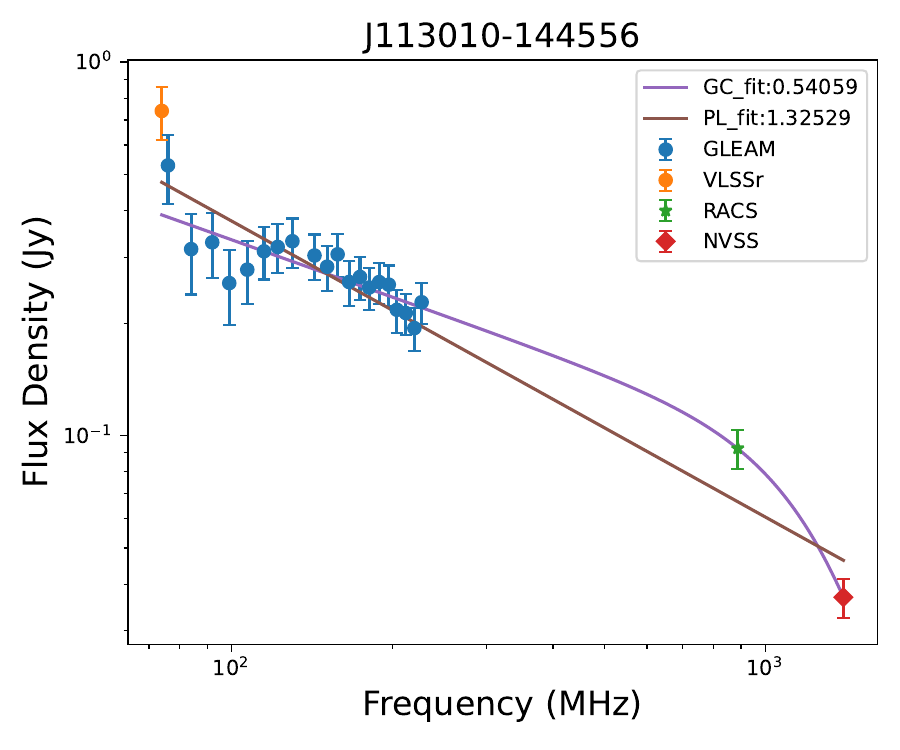}
\caption{Example spectra for sources that are discard in S2d because of $v_{\rm p}>\,600$\,MHz. }
\label{fig:bad}
\end{figure}

We perform the selection of SP sources based on model fitting and F-test results. The applied selection criteria are summarised in Table~\ref{tab:select_routine}, with the following details:

\begin{itemize}
    \item S1) 2,513 sources located at $\delta<-80^{\circ}$ were removed due to the high uncertainty of the flux density and position uncertainty in the GLEAM measurements.
    Furthermore, to ensure the reliable selection of SP sources, it is necessary to guarantee accurate spectra across the entire GLEAM band. 
    Therefore, following the criterion used in \citet[][]{Callingham2017}, sources with fewer than 8 data points in the GLEAM flux density measurements of the signal-to-noise ratio (SNR) $\ge$ 3 are excluded. 
    As a result, additional 58,740 sources were removed from the remaining 304,942 sources. 
    
    We then search for SP sources among the remaining 246,202 sources and further categorize them into the following four samples (S2a to S2d).
    
    \item S2a) The sources in this sample are selected based on the criteria that they have flux density measurements from SUMSS, RACS, or NVSS, with $\alpha \geq 0.1$, and satisfy $\alpha_{_{\rm low}} \geq 0.1$ or meet both of the following conditions simultaneously: $q \geq 0.2$ and $\Delta_q \leq 0.2$.
    Sources with $\alpha \geq 0.1$ are considered potential candidates to have high-frequency peaks in their spectra.
    However, sources that exhibit a flat spectrum within the GLEAM band, but are matched to sources with a higher flux density measured by SUMSS, RACS, or NVSS surveys, can also yield a spectral index of $\alpha>0.1$. 
    We are inclined to attribute such a spectrum to erroneous matching.
    To eliminate possible misselection with $\alpha \ge0.1$, the criterion $\alpha_{_{\rm low}}\ge0.1$ or that satisfies both $q\ge0.2$ and $\Delta_q\le0.2$ is used to guarantee an increasing trend within the GLEAM band or a noticeable upward curvature. 
    A total of 810 sources were selected, out of which 293 were identified in the Callingham's sample. Additionally, 14 sources in this sample also appear in the Convex-Sample provided by \citet[][]{Callingham2017}.
    Convex sources exhibiting a spectrum similar to a power law with a $\alpha<0$ within the GLEAM band (refer to Figure 10 in \citealp[][]{Callingham2017}) are excluded by our criteria due to the lack of sufficient data points outside the GLEAM frequency band to validate such convex spectra. 
    Furthermore, these sources do not significantly impact our subsequent research on the influence of SP sources in the 50--200\,MHz range, as they represent a minor proportion and their spectra within the GLEAM band closely follow a power law.
    In the first row of Figure~\ref{candis}, we present two examples of sources from this sample that were not present in Callingham's sample.

    \item S2b) Sources included in this sample meet the criteria of having $V_{\rm p}\ge 130\,$MHz and a p-value $\le0.005$. These sources exhibit spectra similar to those identified as high-frequency samples in \citet[][]{Callingham2017}. However, unlike their approach of using $\alpha_{_{\rm low}}\ge 0.1$ to determine the presence of a high-frequency peak, we use $V_{\rm p}\ge 130\,$MHz as the criterion. This is because the power-law (PL) model is not accurate enough for fitting SP sources, especially those with low flux density, leading to significant errors in estimating $\alpha_{_{\rm low}}$. 
    The selection of 130\,MHz as the threshold is motivated by the observed correlation between the $V_{\rm p}$ and $\alpha_{\text{low}}$ values of sources in the Callingham's sample. Specifically, we found that the minimum $V_{\rm p}$ value among Callingham's sample sources with $\alpha_{_{\rm low}}\ge 0.1$ is approximately 130\,MHz.
    To ensure the reliability of the peak frequency, we further exclude sources with a p-value $>$ 0.005. 
    With these criteria, a total of 2,745 sources are included in this sample. Among them, 971 sources are also present in the Callingham's sample, and 1 source is found in the Convex-Sample.
    The second row of Figure~\ref{candis} shows the example spectra of two candidates from this sample.

    \item S2c) For sources with lower $V_{p}$ but $V_{p}\ge 72\,{\rm MHz}$, following the selection criteria for low-frequency peaked-spectrum sources proposed by \citet[][]{Callingham2017}, we employ the criterion of $q\le-0.2$ and $\Delta_q$ less than the values defined by Equations 5a and 5b in \citet[][]{Callingham2017} to identify sources that exhibit a noticeable curvature. 
    Because the peaks of these sources appear at low frequencies, limited spectral data are available to assess their authenticity. Therefore, we also require that the p-value be $\le$ 0.005 for the sources in this sample. 
    This selection process yields a total of 792 sources, with 340 sources also found in the Callingham's sample, and 19 sources that overlap with the sample of sources exhibiting peaks below 72\,MHz as provided by \citet[][]{Callingham2017}. The third row of Figure~\ref{candis} shows two newly detected candidates from this sample.
    \item S2d) In this sample, we select sources with $q\le -0.5$, $(q+\Delta_q)\le-0.1$, and a p-value $\le$ 0.005. This sample enriches our collection of peak-spectrum sources by including those that exhibit a significant curve but have $V_p$ < 72\,MHz or $\Delta_q>0.2$. The criteria $(q+\Delta_q)\le-0.1$ ensure a significant curvature in their spectra within the GLEAM frequency band with the errors $\Delta q$ considered at the same time. 
    Since the sources in this sample predominantly have a low-frequency peak, we use the same p-value threshold for this sample, i.e., p-value $\le$ 0.005.
    We also stipulate that $v_{\rm p}$ must be less than 600\,MHz. This criterion leads to the exclusion of 108 sources, as illustrated by the example spectrum shown in Figure~\ref{fig:bad}. 
    This approach yields a sample of 1,174 sources, with 519 of them also included in the Callingham's sample. 
    We present the example spectra of two candidates from this sample in the last row of Figure~\ref{candis}.
\end{itemize}

\begin{figure}
\centering
\includegraphics[width=0.98\hsize]{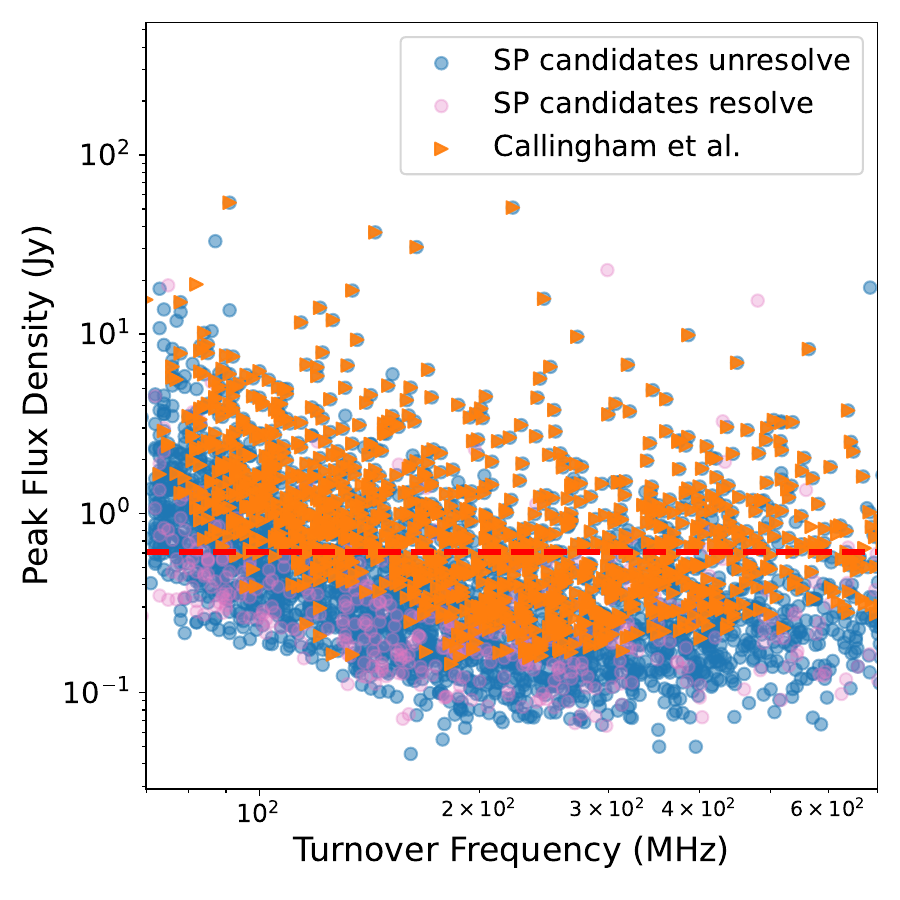}
\caption{The distribution of spectral turnover frequencies and peak flux densities for the sources from GLEAM catalogue where a spectral turnover has been detected. The blue and pink dots represent unresolved and resolved sources in our sample, respectively, while the orange triangles represent the results from Callingham's sample. The red-dashed line indicates the peak flux density at which the total SP sample is considered complete, approximately 0.6\,Jy. }
\label{Sp_Vp}
\end{figure}

\begin{figure}
\centering
\includegraphics[width=8.5cm, angle=0]{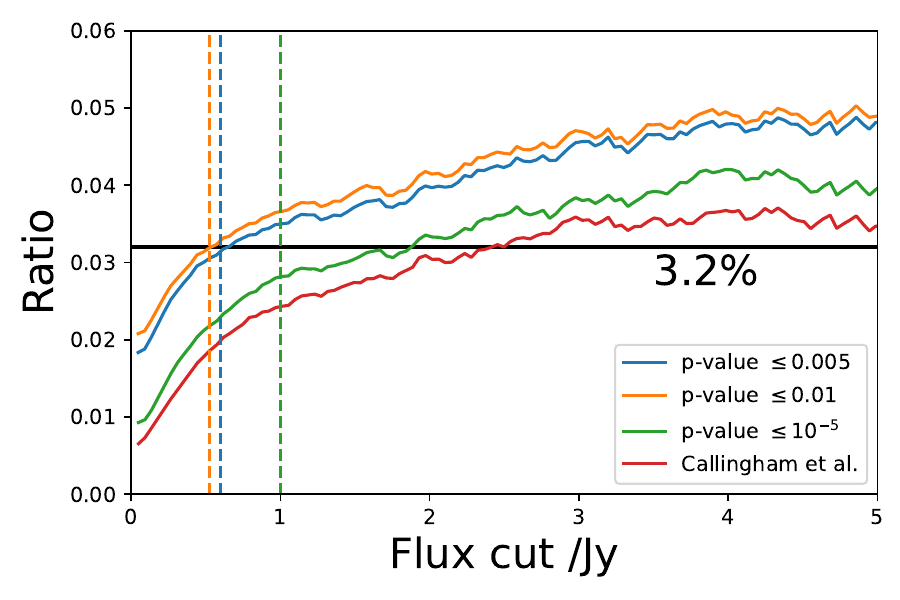}
\caption{The ratio of SP sources to all sources in the GLEAM catalogue. The curves illustrate the fraction $\rm N_{SP, {\it S}_{ 151\,MHz}\ge{\it S}_{cut}}/N_{GLEAM, {\it S}_{151\,MHz}\ge{\it S}_{cut}}$  as a function of flux density cutoff. The colours blue, orange, green, and red represent the results obtained from the samples of SP sources selected with p-value $\le 0.005$, p-value $\le 0.01$, and p-value $\le 1\times10^{-5}$, respectively. The vertical dashed lines of corresponding colours indicate $S_{\rm comp}$, above which they are considered complete. The red vertical dashed line is overlapped by the green one}
\label{ratio}
\end{figure}

We obtained a total of 4,423 SP candidates in these four samples. Note that these four samples are not mutually exclusive.
Among which 496 sources have a $ab/(a_{\rm psf} b_{\rm psf}) \le 1.1$, which are considered as resolved sources \citet[][]{Callingham2017}, where $a$, $b$, $a_{\rm psf}$ and $b_{\rm psf}$ are the semi-major and semi-minor axes of a source and the point spread function, respectively. 
Our sample includes 3,000 sources that were not present in the Callingham's sample from \citep{Callingham2017}. 
Among these, 1,102 sources have flux densities of $S_{\rm wide}<$ 0.16\,Jy, which were not considered in the Callingham's sample. 
Here, $S_{\rm wide}$ represents the flux density in the wideband image. Furthermore, there are an additional 313 resolved sources that were not included in the Callingham's sample due to their choice of exclusion of resolved sources in the study. 
The sources shown in the upper two rows of Figure~\ref{candis} are examples of resolved sources.

We then assess the completeness of our sample by examining the distribution of the peak frequency ($V_{\rm p}$) and the peak flux density ($S_{\rm peak}$) for the candidates for the peaked spectrum. 
As demonstrated by \citet[][]{Callingham2017}, the sample can be considered reasonably complete above a certain flux density, denoted as $S_{\rm comp}$, once the biases introduced by the selection criteria on flux density and frequency are eliminated.
In Figure~\ref{Sp_Vp}, we present the distributions of $V_{\rm p}$ over $S_{\rm peak}$ for both our sample of SP sources and the Callingham's sample. It is evident that an increase in $V_{\rm p}$ results in the identification of more sources with smaller $S_{\rm peak}$ values, consistent with the findings of the Callingham's sample. This suggests that identifying SP sources with lower $V_{\rm peak}$ relies on high signal-to-noise statistics. However, it is worth noting that our sample contains more sources with lower $S_{\rm peak}$ compared to the Callingham's sample. In particular, for peaked spectrum sources with $V_{\rm p}\in$ [72, 80]\,MHz, the lower limit of the 1$\sigma$ distribution of $S_{\rm peak}$ is approximately 0.6\,Jy. This means that for sources with $S_{\rm peak}$ greater than 0.6\,Jy, the dependence of our identification method on $V_{\rm p}$ disappears. Therefore, we can infer that our sample can be considered complete above 0.6\,Jy. 
We further investigate the ratio of the SP source ($\rm N_{SP}$) to all sources after S1 ($\rm N_{GLEAM}$) under varying flux density thresholds at 151\,MHz. 
Applying a series of flux density cuts, we calculate the ratio $\rm N_{SP, {\it S}_{151\,MHz}\ge{\it S}{cut}}/N_{GLEAM, {\it S}_{151\,MHz}\ge{\it S}{cut}}$ (denoted as $\rm N_{SP}/N_{GLEAM}$), which is represented by the blue curve in Figure~\ref{ratio}. 
The ratio at $S_{\rm cut}= 0.6$\,Jy (dashed vertical line) is approximately 3.2 per cent.

In Figure~\ref{ratio}, we have also plotted the results from the Callingham's sample, represented by the red curve.
The Callingham's sample is considered complete for sources with $S_{\rm peak}$ > 1 Jy, as indicated by the vertical red dashed line in Figure~\ref{ratio} (covered by the green dashed line).
Compared to Callingham's sample, there are an additional 309 sources with $S_{\rm peak} > 1$\,Jy in our sample. Among these, 42 sources are resolved or have a low flux density ($S_{\rm wide} < 0.16$\,Jy), which were not considered in the Callingham's sample.
Out of these 42 sources, 33 are resolved sources, indicating that the majority of the 309 sources are unresolved. Additionally, 280 of them have $\nu_{\text{peak}} > 72$ MHz, but only 63 have $\nu_{\text{peak}} > 130$ MHz, suggesting that most of them have a peak very close to 72 MHz. Since the majority of these sources are concentrated in the low $v_{\rm peak}$ region, therefore, the differences between the two samples are only evident in this specific region, suggesting overall good agreement between the two samples in Figure~\ref{Sp_Vp}.
In Callingham's sample, resolved or low flux density sources are eliminated to ensure the reliability of identified peaked-spectrum sources. However, our sample selection of 4,423 SP sources did not impose such restrictions, resulting in our sample not having the same level of reliability as Callingham's sample. If sources are resolved, the variation in resolution among different observational devices can lead to incorrect spectra. For example, heavily resolved sources in GLEAM may not be fully captured by NVSS, resulting in inaccurate spectra.
Our objective is to examine the influence of spectra with peaks on EoR detection, rather than solely identifying reliable astrophysical peaked-spectrum sources. Therefore, our focus is to identify as many potential sources or factors that could influence EoR detection as possible. In Section~\ref{sect:results}, we test the potential influence of SP sources without eliminating resolved sources. This decision is based on two factors. First, the peaks of the incorrect spectra consistently appear at high frequencies, which have a relatively small influence within the 50--200\,MHz range that we are interested in. Second, EoR detection may also be affected by the erroneous spectra caused by the resolutions of the devices.

\begin{figure}
\centering
\includegraphics[width=8.5cm, angle=0]{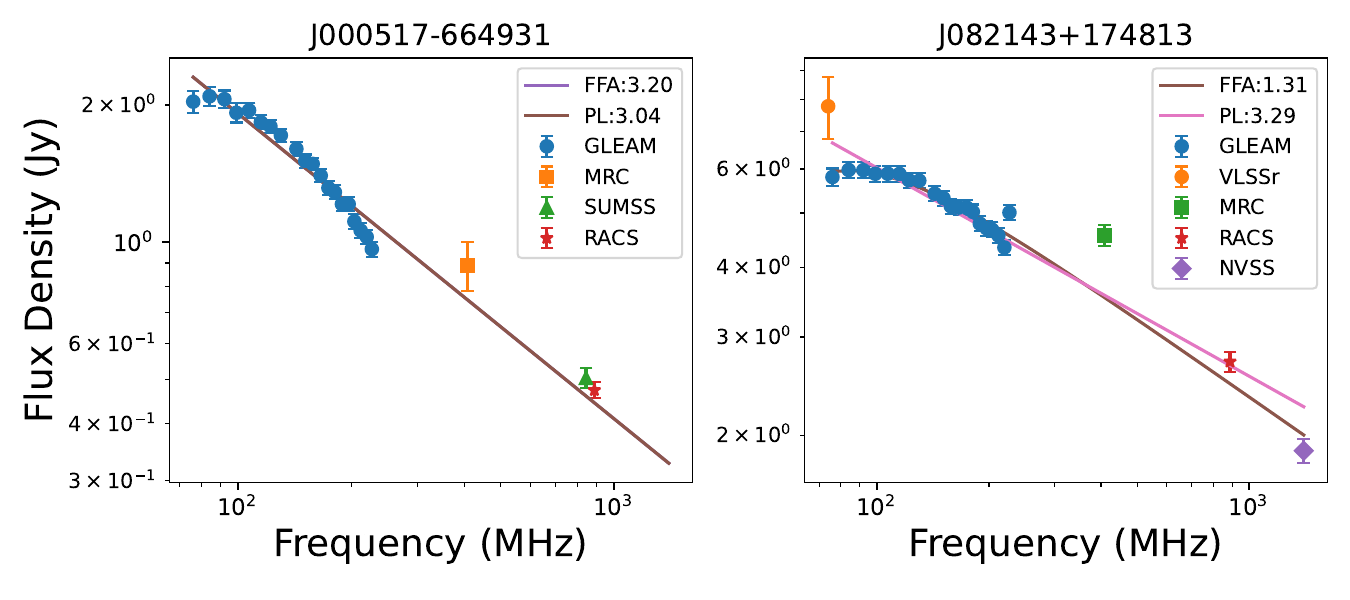}
\caption{Example spectra for sources found in the Callingham's sample but not in our sample. The source in the left panel was excluded due to a large p-value, while the source in the right panel was omitted due to lack of significant curvature.}
\label{cal_lose}
\end{figure}

Among the sources in the Callingham's sample, 60 sources do not meet our criteria for being classified as SP sources. Out of these, 49 sources are excluded due to having high p-values. An example spectrum of such sources is illustrated in the left panel of Figure~\ref{cal_lose}. The remaining sources are not selected due to the lack of significant curvature, and an example spectrum of such sources is displayed in the right panel of Figure~\ref{cal_lose}. Since the composition of the SP source samples is highly dependent on the choice of p-value thresholds, we examine how the sample of SP sources varies with different p-value thresholds, which in turn affects the ratio $\rm N_{SP}/N_{GLEAM}$. This ratio will be used for further investigation of its impact on EoR detection. We vary the p-value threshold from $1\times10^{-5}$ to $0.01$ when selecting the S2b, S2c, and S2d samples. As the p-value threshold increases, the number of candidates ($\rm N_{SP}$) in the samples increases progressively from 2,222 to 5,030. Furthermore, with an increase in the p-value threshold, the number of sources appearing in the 1,483 samples but not selected as SP sources (denoted as $\rm N_{lose}$) decreases from 284 to 52.

\begin{figure}
\centering
\includegraphics[width=0.98\hsize]{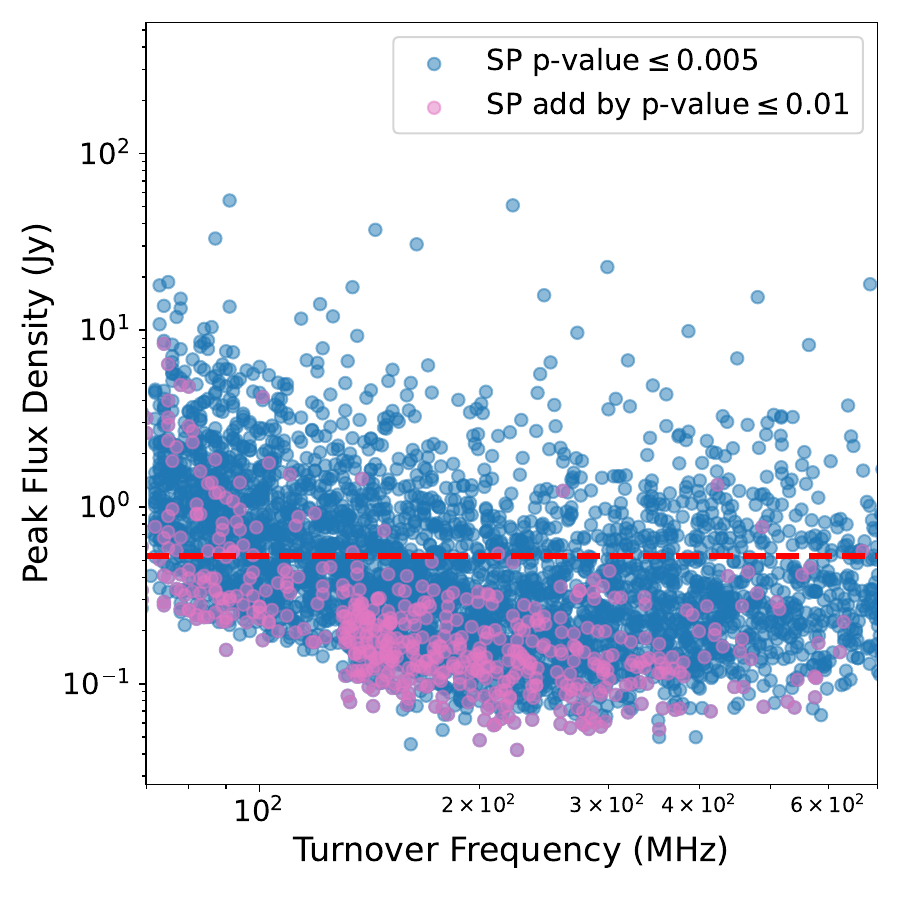}
\caption{Same as Figure~\ref{Sp_Vp}, but for the SP sample selected with p-values $\leq$ 0.01 (for S2b, S2c and S2d). The blue dots represent SP sources selected with p-value$\leq0.005$ (shown in Figure 3). The additional points found with a p-value threshold of $\leq 0.01$ are represented by the pink dots. The red dashed line represents $S_{\rm comp}$, which is approximately 0.5\,Jy.}
\label{Sp_Vp_2}
\end{figure}

Subsequently, we examine how $S_{\rm comp}$ varies with different p-value thresholds. As the p-value threshold increases, the candidates become complete at a lower $S_{\rm peak}$, resulting in $S_{\rm comp}$ decreasing from 0.9 Jy to 0.5 Jy. The distribution of $V_{\rm p}$ and $S_{\rm peak}$ for the SP sources selected with p-values $\le$ 0.01, which we used to determine $S_{\rm comp}$, is shown in Figure~\ref{Sp_Vp_2}.
The ratio $\rm N_{SP}/N_{GLEAM}$ of the samples selected by p-value $\le 0.01$ and p-value $\le1\times10^{-5}$ as a function of $S_{\rm cut}$ values is illustrated in Figure~\ref{ratio} using the orange line and green line respectively. 
The corresponding $S_{\rm comp}$ for these samples are marked by dashed vertical lines with the same colours as the ratio lines. 
The less strict p-value will result in a sample with more candidates, causing $\rm N_{SP}/N_{GLEAM}$ at a fixed $S_{\rm cut}$ to increase. 
However, the samples selected with less strict p-value can be complete at a lower $S_{\rm comp}$, and the values of $\rm N_{SP}/N_{GLEAM}$ at a lower $S_{\rm cut}$ are smaller. 
Therefore, the ratio of $\rm N_{SP}/N_{GLEAM}$ counted for the sources brighter than the limit of flux density $S_{\rm comp}$ can remain relatively stable. 
We have observed that the ratio of $\rm N_{SP}/N_{GLEAM}$ in the resulting samples, obtained with different p-value thresholds, varies from 2.7 to 3.2 per cent at the corresponding $S_{\rm comp}$ cutoff. Furthermore, when considering only the unresolved sources, the results remain consistent with those obtained from the sample that includes resolved sources, with a similar variation ranging from 2.7 to 3.2 per cent. The difference in ratios between the sample with resolved sources and the sample without resolved sources falls within a range of 0.07 per cent. Since our main objective is to investigate the potential impact of SP sources, we use this maximum ratio, i.e., 3.2 per cent, for further investigation of the impact on EoR detection.

\section{Influence of Spectral Unsmoothness in EoR Signal Detection}
\label{sect:results}
In this section, we will set up mock observations that include EoR signals and extragalactic radio sources, a fraction of which are SP sources. We will investigate the impact of SP sources on EoR signal extraction experiments based on these mock observations.

\subsection{Simulations}
\label{subsect:simulations}
 We simulate the observation of the EoR signal and foreground extragalactic sources from a sky area of $10^{\circ} \times 10^{\circ}$. We use the Gaussian profile for the simulated synthesised beam. The angular resolution is 4.7\,arcsec at 200\,MHz and 18.9\,arcsec at 50\,MHz, corresponding to the observation baseline of 80 km. Since we concentrate on studying the influence of spectral unsmoothness of extragalactic sources, the Milky Way emission is not included in the simulations. 

The simulated EoR signal is based on data from the {\it Evolution Of 21 cm Structure} project\footnote{Evolution Of 21 cm Structure project: \url{http://homepage.sns.it/mesinger/EOS.html}} released in 2016. 
The project uses \textsc{21cmFAST} to simulate the cosmic reionization process from redshift 86.5 to 5.0 inside a large data cube with each side of $\rm 1.6\,comoving\,Gpc$ (1024 cells). 
Using the tools provided by the \textsc{fg21sim} package\footnote{FG21sim: \url{https://github.com/liweitianux/fg21sim}}, we extract the image slices at 50--200\,MHz from the light-cone cubes of “faint galaxies” case, and then tile and rescale each slice according to the sky coverage and pixel size of our simulation.



Four types of sources are employed as the simulated extragalactic point sources: (1) star-forming and starburst galaxies, (2) radio-quiet AGNs, (3) Fanaroff–Riley type I and type II AGNs , and (4) peaked-spectrum sources which mainly consist of GPS AGNs and CSS AGNs. 
The first three types are simulated based on the semiempirical extragalactic data $S^3$ (hereafter $S^3$ data) simulated by \citet{Wilman2008}. 
In the $S^3$ sample, star-forming and starburst galaxies account for approximately 79.0 per cent, radio-quiet AGNs make up 13.3 per cent, and Fanaroff–Riley type I and type II AGNs constitute 7.7 per cent. 
When simulating these types of galaxies, the spectra of star-forming, starburst galaxies and radio-quiet AGNs follow a power-law spectrum with a spectral index of -0.7. Additionally, the extended lobe emission spectrum of Fanaroff–Riley type I and type II AGNs exhibit a power-law spectrum characterized by a spectral index of -0.75. Meanwhile, the spectrum of core emissions adhere to the formula described by:
\begin{equation}
    {\rm log}S_{v}^{\rm core}= a_0 + a_1 {\rm log}(\frac{v}{\rm GHz}) + a_2 {\rm log}^2(\frac{v}{\rm GHz}),
\end{equation}
here $v$ indicates the frequency, $a_1=0.07$, $a_2=-0.29$, and the values of $a_0$ are calculated from the flux densities provided by the $S^3$ simulation.
More details can be found in \citet{Wang2010} and references therein. 
The spectra of SP sources are simulated based on spectral properties of the samples we get in Section \ref{sect:Candidates}.

The construction of mock observations is detailed below:
\begin{itemize}
    \item 1) We mask exceptionally bright sources with a flux density exceeding 200\,Jy at 151\,MHz ($S_{\rm 151MHz}>200\,{\rm Jy}$) from the $S^3$ simulation. Such exceptionally bright sources are rare, particularly within a limited sky area. As a result, we exclude an extremely bright source with a flux density of $S_{\rm 151MHz}=943\,{\rm Jy}$. With the removal of this source, the average brightness temperature of the foreground at 50\,MHz is reduced by a factor of approximately 50. In the data processing of real observations, such bright sources will be removed before the data is used to extract EoR signals. 
    \item 2-A) We constructed a counterpart mock comprised entirely of power law sources ( hereafter referred to as `3.2\%-PL') for comparison. We randomly selected 3.2 per cent of the sources and adjusted their spectral indices to match the spectral index distribution derived from sources in the GLEAM catalogue. This ensures that the spectral indices of sources in the power-law simulation are not limited to a few specific values \citep[see][]{Wang2010}.
    
    \item 2-B) For the foreground mock that incorporates SP sources (hereafter 3.2\%-SP), we substitute the 3.2 per cent $S^3$ sources with SP sources. The spectra of these SP sources are generated based on the spectra of the SP sources we identified in Section~\ref{sect:Candidates}, but we rescale the amplitude of the spectra to ensure that the sum of $S_v$ over all frequencies (50--200\,MHz) remains the same as the 3.2\%-PL simulation.
    
    \item 3) We overlay the simulated foreground sky maps with 21-cm signal maps at the same frequencies, and then convolve the combined maps with the simulated Gaussian synthesised beams at corresponding frequencies. 
    
\end{itemize}

\begin{figure*}
\centering
\includegraphics[width=0.98\hsize]{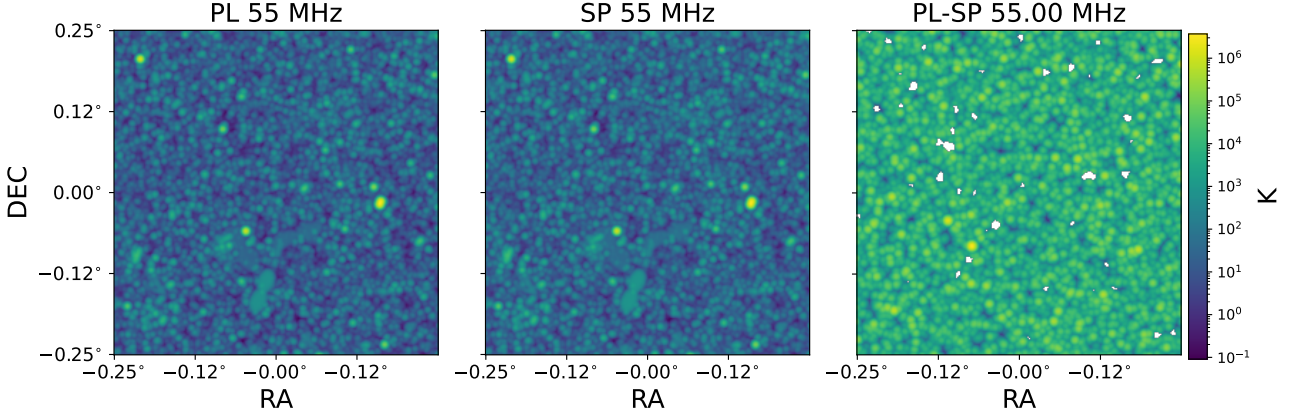}
\hfill
\caption{Simulated $0.5^{\circ} \times 0.5^{\circ}$ sky maps without (left panel) and with 3.2 per cent (middle panel) SP sources at 55\,MHz, as well as the difference between these two simulated maps (right panel).}
\label{sim_sources}

\end{figure*}

\begin{figure}
\centering
\includegraphics[width=0.98\hsize]{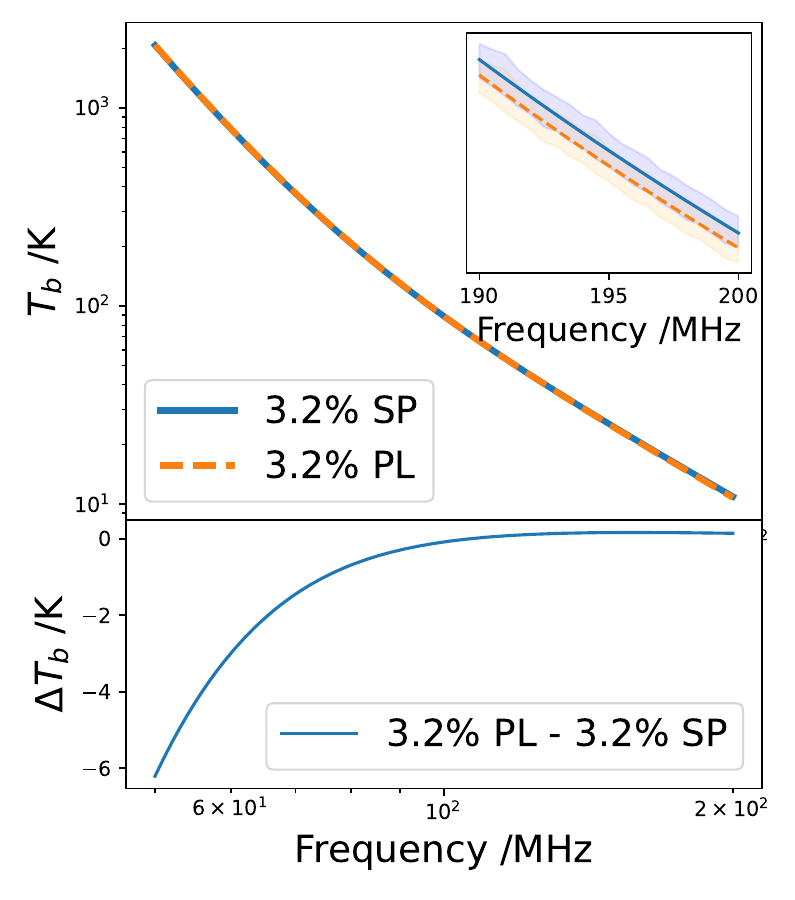}
\caption{The averaged brightness temperature ($T_b$) of simulations with and without SP sources, as well as the difference between the $T_b$ values of two simulations (bottom part). The dashed orange line indicates the $T_b$ of the simulation with 3.2 per cent SP sources (3.2\%-SP), while the solid blue line refers to the $T_b$ of the simulation without SP sources (3.2\%-PL). The upper right inset provides a zoomed-in view of the frequency range from 190\,MHz to 200\,MHz. The shaded areas around the lines represent the 1$\sigma$ errors of the measurements, which were estimated using 200 bootstrap resamplings.}
\label{fig:global}
\end{figure}

In Figure~\ref{sim_sources}, we present the $0.5^\circ \times 0.5^\circ$ simulated sky maps and their differences at 100\,MHz. The average brightness temperature ($T_b$) across the entire $10^\circ \times 10^\circ$ sky image slice, as a function of frequency, is shown in Figure~\ref{fig:global}.
These two simulations show remarkable similarities in both the images and the averaged brightness temperature. The slight differences are primarily observed at the extreme frequencies, which result from the constraint of maintaining the same total flux density across all frequencies in both simulations.
The largest difference in the average brightness values is 6.2\,K at 50\,MHz, while at 200\,MHz, the difference is 0.14\,K (approximately 1.3 per cent). These differences are within the $1-\sigma$ error estimated through 200 bootstrap resamplings \citep[][]{Bootstrap1976}. 





\subsection{Statistical analysis for influence of SP sources}
\label{statistic}

\subsubsection{Foreground removal with polynomial fitting}
\label{sec:poly-fit}
We start by using polynomial fitting to remove foregrounds. Since our primary focus is not on studying foreground removal techniques, we directly apply a pixel-by-pixel algorithm to subtract the simulated extragalactic sources.
We test the polynomial fitting in the fitting frequency bandwidths 15\,MHz and 30\,MHz. 
The polynomial orders are adjusted to balance between minimising foreground residuals and preserving a significant portion of the EoR signal.
The set of $\rm n^{th}$-order polynomial are defined as $T_b= \sum_{i=1}^{n}a_i v^{i}$, where $a_i$ indicates the coefficients of polynomial. 
For a frequency bandwidth of 15\,MHz, we use $7^{\rm th}$-order for spectra of 50--65\,MHz and $5^{\rm th}$-order for each 15\,MHz width spectrum in the range 65--200\,MHz. 
For a wider fitting bandwidth of 30\,MHz with higher-order polynomial, $9^{\rm th}$-order ($\le 80$\,MHz) and $7^{\rm th}$-order (80--200\,MHz).
We use higher-order polynomials for low-frequency data to remove the stronger low-frequency foreground contamination more thoroughly.

\begin{figure*}
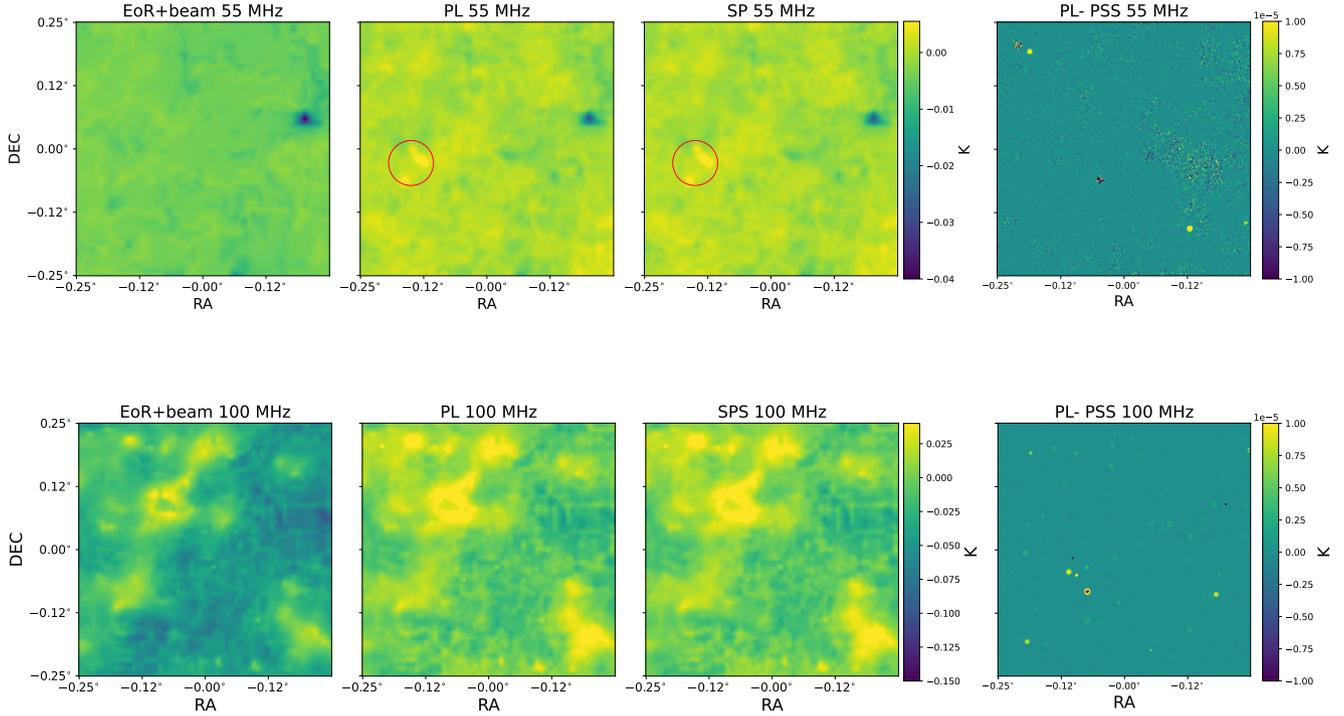

\centering
\includegraphics[width=0.98\hsize]{fig_new/res_image\_55MHz.pdf}
\includegraphics[width=0.98\hsize]{fig_new/res_image\_100MHz.pdf}
\caption{Simulated $0.5^{\circ} \times 0.5^{\circ}$ EoR signals (left panels, beam effects are included) and residuals after foreground subtraction with a $\rm 9^{th}$-order polynomial (bandwidth 15\,MHz) at 55\,MHz (upper panels) and a $\rm 5^{th}$-order polynomial (bandwidth 15\,MHz) at 100\,MHz (bottom panels). From left to right, the panels show the simulated EoR signal, the residuals of the 3.2\%-PL simulation, the residuals of the 3.2\%-SP simulation, and the differences between the residuals of the two simulations, respectively. The three columns on the left share common colorbars displayed to the right of the third column. The red circles indicate obvious foreground residuals.}
\label{res_fig}
\end{figure*}

Figure~\ref{res_fig} shows the foreground-removed sky maps from aforementioned mock observations in which the EoR signal, extragalactic sources, and beam effects are included. After the foreground is removed, the structures of the input EoR signal are recovered.
Note that if not specified, the `residual' in this paper indicates residual after foreground removal, which encompasses both the foreground residual and the EoR signal.
At 55\,MHz, the foregrounds are removed with $9^{\rm th}$-order polynomial (15\,MHz bandwidth) instead of $7^{\rm th}$-order polynomial (15\,MHz bandwidth), since $7^{\rm th}$-order polynomial (15\,MHz bandwidth) at such low frequency would leave obvious foreground residuals. 
 However, the temperature of residuals is higher than that of the input EoR signal generally, suggesting that the foreground is not removed completely.
In the low-frequency residual maps at 55\,MHz, some structures from the residuals of the foreground are obvious in both the 3.2\%-PL residual and the 3.2\%-SP residual, as illustrated by the red circles in Figure~\ref{res_fig}.
The fact that these structures vanish in the differences between residuals (right panel) implies that they are foreground residuals that failed to be removed, rather than extra structure introduced by SP sources. The residual maps of the two mock observations are quite similar to each other, with differences between the median of residual maps at the $1\times10^{-7}$ K level.

\begin{figure*}
\centering
\includegraphics[width=0.98\hsize]{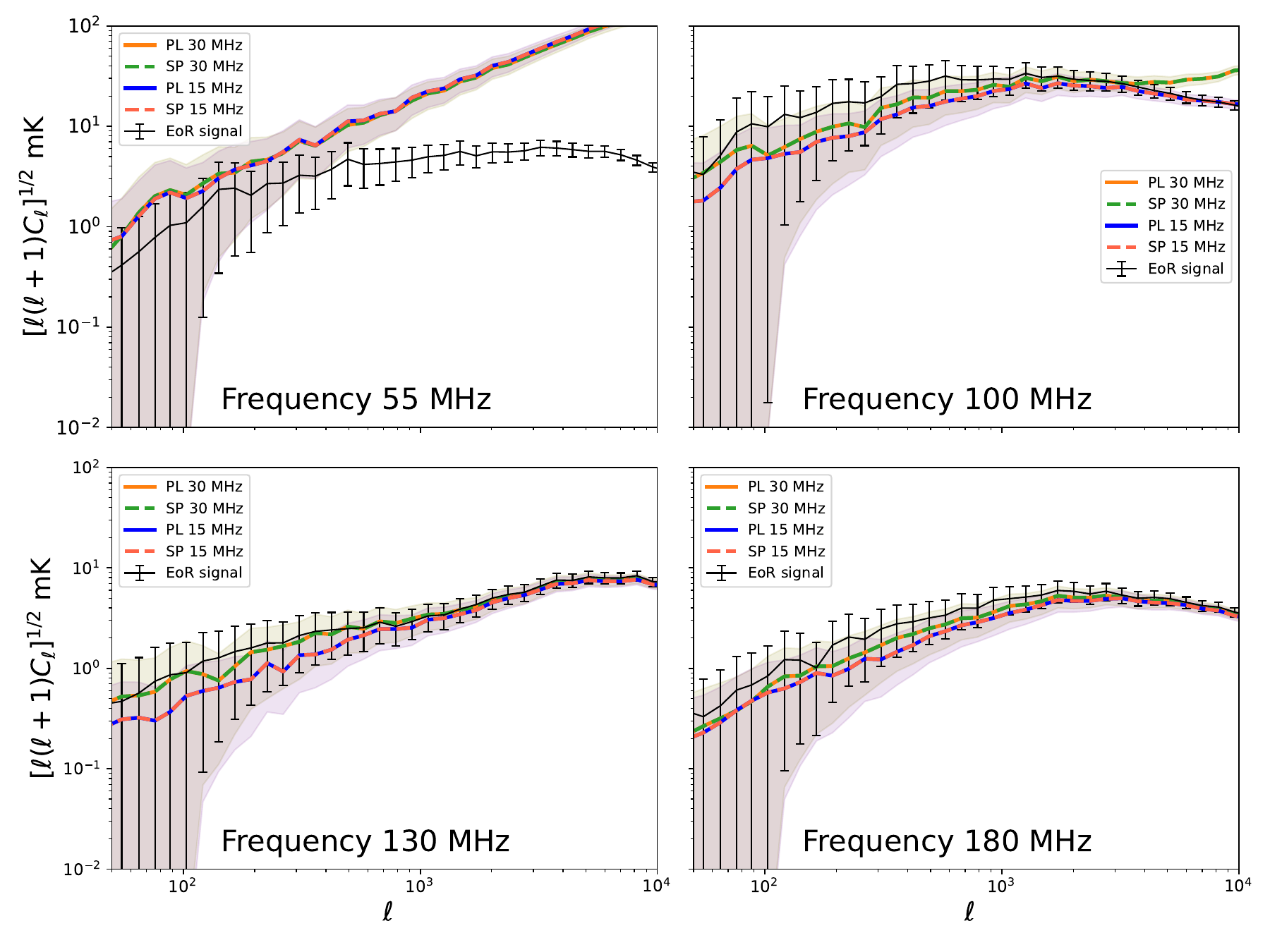}
\caption{The angular power spectra of EoR signal (black), and residuals (15\,MHz: red dotted and blue solid lines; 30\,MHz: golden solid and green dotted lines) after foregrounds subtraction with different fitting bandwidth and polynomial order. The spectra from left to right and from top to bottom correspond to frequencies of 55\,MHZ, 100\,MHz, 130\,MHz and 180\,MHz, respectively. 
At all frequencies, the angular power spectra of residuals from the 3.2\%-PL simulations (solid lines) and the 3.2\%-SP simulations (dotted lines) overlap when the fitting bandwidth and polynomial order are the same. Error bars of the EoR angular power spectrum and shaded regions filled with the same colour around individual spectral lines are used to represent the uncertainty (estimated by Eq.~\ref{eq:delta_Cl}) associated with the respective spectral lines. }
\label{ps_1D}
\end{figure*}
We then assess the influence of SP sources on the angular power spectrum. Figure~\ref{ps_1D} presents the angular power spectra of the residuals and the EoR signal at 55\,MHz, 100\,MHz, 130\,MHz and 180\,MHz. The results suggest that although the power of foregrounds is 4$\sim$6 orders of magnitude higher than that of the EoR signals, the residuals after foreground removal are close to the EoR signal. We calculate the angular power spectra with \textsc{Healpy}\footnote{\url{https://healpy.readthedocs.io/en/latest/tutorial.html}}. The variance of the power spectra $\Delta C_{\ell}$ is estimated with 
\begin{equation}
    \Delta C_{\ell} = [2/(2\ell+1)f_{\rm sky}]^{1/2}(C_{\ell}+N_{\ell})
\label{eq:delta_Cl}
\end{equation}
where $N_{\ell}$ is the noise power spectrum defined by $N_{\ell} = (wf_{\rm sky})^{-1}e^{\theta_b^2\ell(\ell+1)}$, and $\theta_{b}$ indicates the width of the half-power beam of the synthesised beam. The contribution of white noise, denoted as $w^{-1}$, is expressed as $w^{-1}= 4\pi\sigma_{\rm pix}^2/N_{\rm pix}$, where $\sigma_{\rm pix}$ and $N_{\rm pix}$ are the pixel noise and the total number of pixels, respectively. 
$\sigma_{\rm pix}$ can be estimated by $\sigma_{\rm pix}=T_{\rm sys}/\eta \sqrt{2N\Delta vt}$ in radio interferometric measurements. 
In this work, we assume a system temperature $T_{\rm sys}$ is 200~K, an efficiency factor of the telescope $\eta = 0.8$, a total number of independent baselines $N=(512\times511)/2$, which is the same as the SKA-low, a frequency bandwidth of $\Delta v=0.5$ MHz, an observing time $t = 1$ year, and The $f_{\rm sky}= (\pi\times100)/129600$, where 100 represents the area of the sky coverage we simulated, expressed in square degrees.

For all the frequencies that we tested, the residuals of the two simulations are almost the same. The differences between the power spectra of the residuals are much lower than the $\Delta C_{\ell}$ (error bars). 

We check the difference between the residuals of the 3.2\%-PL and 3.2\%-SP simulations with $\ell$ in the range of $[50, 10^4]$ and compare it with the EoR signal. 

At high frequencies, such as 100\,MHz, 130\,MHz, and 180\,MHz, most of the foreground can be removed by polynomial fitting. Differences for $C_{\ell}$ between residuals ($<10^{-6}$ mK, $<10^{-8}$ mK, $<10^{-10}$ mK at 100, 130, and 180 MHz, respectively) are at least $5 \sim 6$ orders of magnitude lower than  $C_{\ell}$ of the EoR signal.

\begin{figure}
\centering
\includegraphics[width=0.98\hsize]{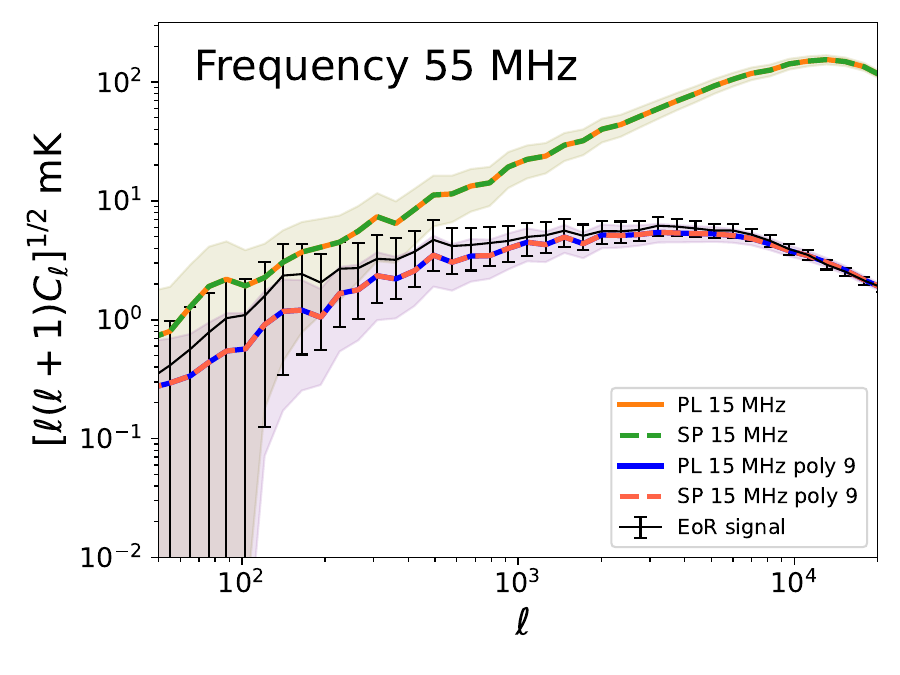}
\caption{Similar to Figure~\ref{ps_1D}, this presents the results obtained when removing foreground with a $\rm 9^{th}$-order polynomial within a 15\,MHz bandwidth spectrum (3.2\%-PL: blue solid line; 3.2\%-SP: red dashed line). For comparison, the golden solid line (3.2\%-PL) and the green dashed line (3.2\%-SP) depict the residuals after foreground removal using a $\rm 7^{th}$-order polynomial with a 15\,MHz bandwidth, which are also shown in Figure~\ref{ps_1D}. }
\label{fig:55MHz-9order}
\end{figure}

For the low-frequency case (55\,MHz, see the upper left panel of Figure~\ref{ps_1D}), foreground residuals are much higher than the EoR signal, especially at small scales of $\ell>\approx 400$. 
It suggests that the method of foreground removal does not work well for the two mock observations at low frequency. 
However, the discrepancy between residuals is less than 5.5 per cent of the EoR signal, which suggests the influence of SP sources on the removal of the foreground is small. 
We refine the foreground subtraction with a new setting of fitting a 15 MHz bandwidth spectrum with a $9^{\rm th}$-order polynomial and show the angular power spectra of the residuals in Figure~\ref{fig:55MHz-9order}. 
Increasing the polynomial order or narrowing the fitting bandwidth leads to a more effective removal of the foreground, particularly at $\ell>\approx$ 1000. 
However, this improvement comes at the cost of some loss of the EoR signal, especially for $\ell<\approx$ 1000. 
In this case, the difference between the residuals of 3.2\%-SP and 3.2\%-PL is less than 0.033 per cent of the EoR signal and less than 0.5 per cent of the difference between the EoR signal and the residuals of 3.2\%-PL. 
The influence of SP sources on the removal of the foreground is still small. 



\begin{figure}
\centering
\includegraphics[width=0.98\hsize]{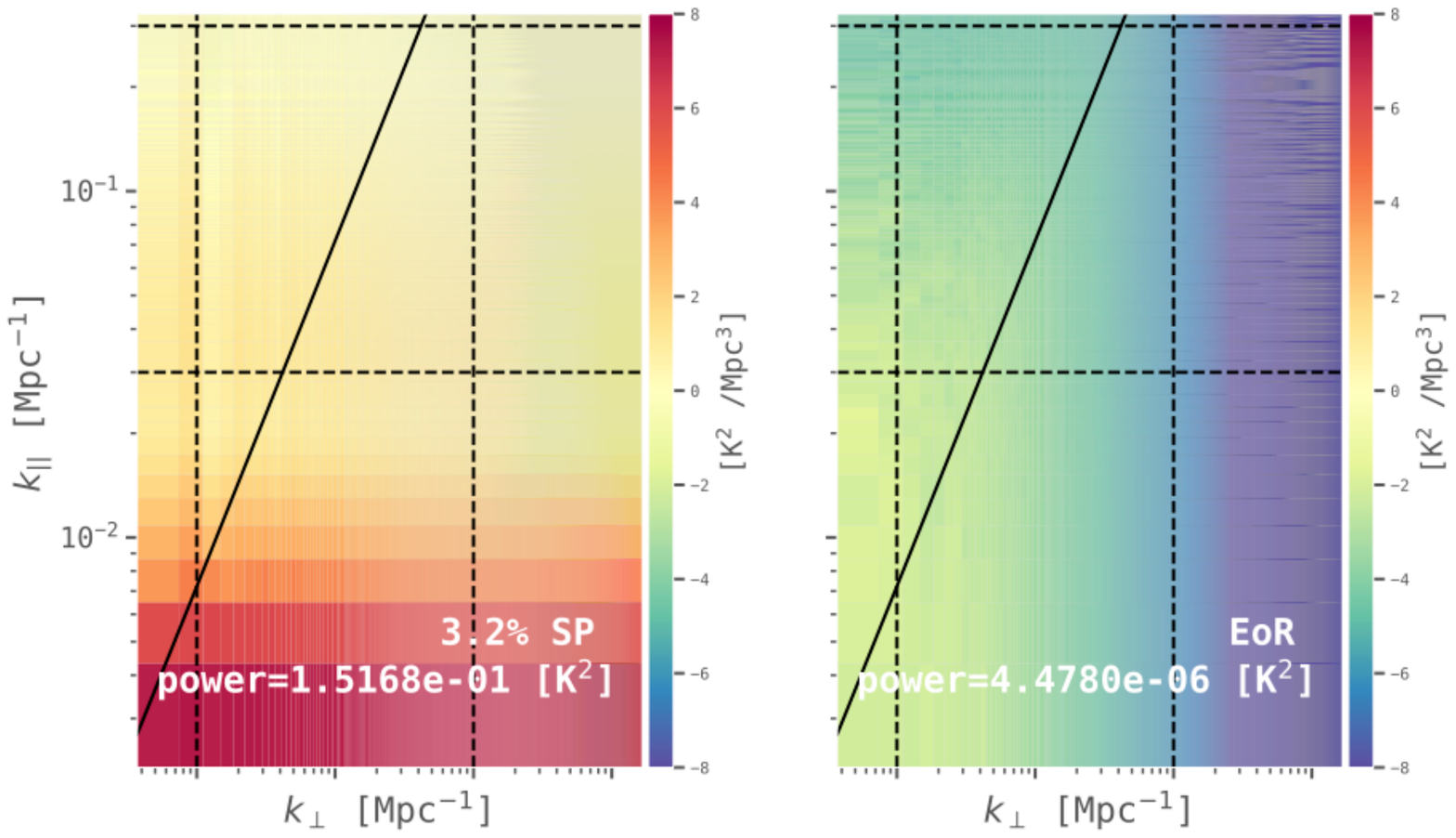}
\caption{The 2D power spectra of EoR signal (right panel) and simulated sky maps (including extragalactic sources and EoR signal, left panel), both are the results of sky map convolved with Gaussian beams. The solid line shows boundary between the EoR window (top left) and the foreground wedge (bottom right). The values of total power in EoR window are shown in the bottom right corner of each 2D power spectrum figure. }
\label{PS_2D_fg_21}
\end{figure}

We then examine the differences in the two-dimensional (2D) power spectra in the $k_{\perp}, k_{\parallel}$ plane, which serves as a valuable diagnostic tool to identify distinct contamination regions. In the $\{k_{\perp}, k_{\parallel}\}$-space, foreground contamination is observed in specific regions known as the "foreground wedge". Above the foreground wedge regions lies the `EoR window', where the upper limit on the 21 cm reionization power spectrum can be measured \citep[][]{Beardsley2016, Barry2019}. We show an example of 2D power spectra of the simulated sky map without foreground subtraction (left panels) and the simulated EoR signal (right panel) in Figure~\ref{PS_2D_fg_21}. The solid lines indicate the boundary between the foreground wedge and the EoR window, described by \citep{Thyagarajan2013}:
\begin{equation}
    k_{\parallel}\ge \frac{H(z)D_M(z)}{(1+z)c}[k_{\perp}sin\theta +\frac{2\pi wf_{21}}{(1+z)D_M(z)B} ]
\label{Eq:EoR_win}
\end{equation}
where frequency bandwidth of simulated image cube $\rm B= 150$\,MHz, $f_{21} = 1420.4$\,MHz indicates the rest-frame frequency of 21 cm signal, $z$ is the redshift corresponding to the central frequency of the image cube, the $H(z)$ and $D_M(z)$ refer to the Hubble parameter and angular diameter distance at redshift $z$ respectively. Additionally, $c$ represents the speed of light and $\theta$ represents the angular distance of foreground sources from the centre of the field. 
The total power within the upper left region, bounded by the black solid line and dashed lines, which satisfies Eq.~\ref{Eq:EoR_win}, $\rm 0.03\, Mpc^{-1}<k_{\parallel}<0.3\,Mpc^{-1}$ and $\rm 0.01\,Mpc^{-1}<k_{\perp}<1\,Mpc^{-1}$, is denoted as $\rm P_{win}$, and the values are displayed in the bottom right corner of each 2D power spectrum figure.

\begin{table*}
\centering
    \begin{tabular}{ccccc}
    \hline
    & poly-order & all in res(K$^2$)  & FG only res(K$^2$)  \\
    \hline
    SP 30\,MHz & 80\,MHz :9; 80--200\,MHz: 7 & $4.3630 \times 10^{-06}$ & $1.0696 \times 10^{-07}$ \\
    PL 30\,MHz & 80\,MHz :9; 80--200\,MHz: 7 & $4.3630 \times 10^{-06}$ & $1.0698 \times 10^{-07}$ \\
    SP 15\,MHz & 65\,MHz :7; 65--200\,MHz: 5 & $3.9397 \times 10^{-06}$ & $2.6064 \times 10^{-07}$ \\
    PL 15\,MHz & 65\,MHz :7; 65--200\,MHz: 5 & $3.9397 \times 10^{-06}$ & $2.6052 \times 10^{-07}$ \\
    \hline
    \end{tabular}
    \caption{Results of foreground removal. For each case, we perform polynomial fitting with two different fitting frequency bandwidths: 30\,MHz and 15\,MHz. The `poly-order' column lists the order of the polynomial we use to fit the spectra. The last two columns display the total powers within the EoR window of the 2D power spectrum. The column labelled `FG only res' presents the results obtained from simulations that include only extragalactic sources, while the column labelled `all in res' shows the results obtained from simulations that include extragalactic sources, the EoR signal, and beam effects.}
    \label{tab:polyfit_res}
\end{table*}

\begin{figure}
\centering
\includegraphics[width=0.98\hsize]{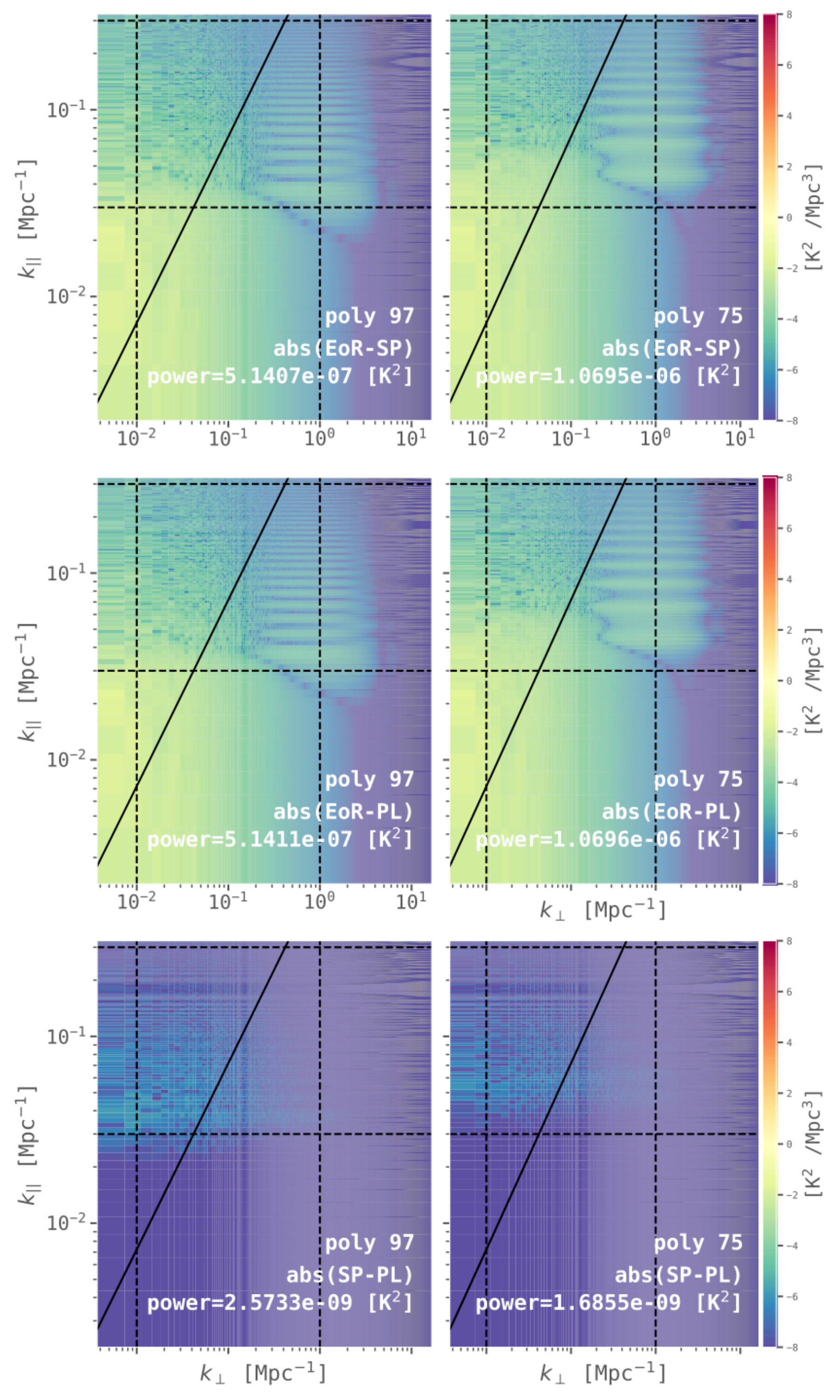}
\caption{The difference between 2D power spectra of EoR signal and residuals after foreground subtraction (upper two rows) and difference between residuals of two simulation (bottom row). The left panels represent results obtained with a fitting bandwidth of 30\,MHz, while the right panels represent results obtained with a fitting bandwidth of 15\,MHz. From top to bottom are differences between EoR signal and residuals of 3.2\%-SP, EoR signal and residuals of 3.2\%-PL, residuals of 3.2\%-SP and 3.2\%-PL, respectively. The three panels on the left share a common colour bars of right panels.}
\label{PS_2D}
\end{figure}

$\rm P_{win}$ of the residual maps, as well as the corresponding foreground removal settings, are summarised in Table~\ref{tab:polyfit_res}. 
The last column (`FG only res') shows results from foreground-only simulation, i.e., EoR signal is not included in the simulation. 
It shows that $\rm P_{win}$ of residuals are only 2--6 per cent of the $\rm P_{win}$ of EoR signal (marked as $\rm P_{win, EoR}$, shown in the lower right corner of the right panel in Figure~\ref{PS_2D_fg_21}).
The penultimate column, labelled `all in res', show the $\rm P_{win}$ of residuals of simulations in which the EoR signal, extragalactic sources, and beam effects are all included. 
In Figure~\ref{PS_2D}, we also present the difference between the 2D power spectrum of the simulated EoR signal and the residuals of the two simulations in the `all in' scenario. 
The absolute difference between the 2D power spectra of X and Y is denoted as $\rm PS_{|X-Y|}$. For example, $\rm PS_{|EoR-SP|}$ represents the absolute difference between the 2D power spectrum of the EoR signal and the residual of the 3.2\%-SP simulation.
Then we use $\rm P_{win, |X-Y|}$ to represent the total power in the EoR window of the corresponding $\rm PS_{|X-Y|}$.
We find that $\rm P_{win, |EoR-PL|}$ and $\rm P_{win, |EoR-SP|}$ deviate by approximately 11 per cent (30\,MHz bandwidth) and 24 per cent (15\,MHz bandwidth) from $\rm P_{win, EoR}$.

However, $\rm P_{win, |SP-PL|}$ is significantly smaller than $\rm P_{win, |EoR-PL|}$ (or $\rm P_{win, |EoR-SP|}$), accounting for only about 0.5 per cent (30\,MHz bandwidth) and 0.16 per cent (15\,MHz bandwidth) of $\rm P_{win, |EoR-PL|}$, and only 0.06 per cent (30\,MHz bandwidth) and 0.04 per cent (15\,MHz bandwidth) of $\rm P_{win, EoR}$. It implies that the additional power in the EoR window introduced by SP sources can be negligible, as they are approximately 2--3 orders of magnitude lower than the residuals from the entire extragalactic point sources with the current method of foreground removal.
If alternative settings or different foreground removal methods are used, the results may be different. To explore the potential influence of foreground removal methods, we also evaluated the results of using two additional foreground subtraction methods, PCA and FastICA.

\subsubsection{Foreground Removal With PCA And FastICA}
We examine the difference in residuals between the mock observations of 3. 2\%-PL and 3. 2\%-SP after removing the foregrounds using PCA and FastICA\footnote{\url{https://scikit-learn.org/}}. 
For both methods, we use 3, 9, and 15 principal components to identify the foregrounds, denoted as $\rm N_{FG}=[3, 9, 15]$, and then subtract them to compare the residuals with the input EoR signal.
\begin{table}
    \centering
    \begin{tabular}{lccc}
    \hline
    \multirow{2}*{Method and $\rm N_{FG}$}& \multicolumn{3}{c}{$\rm P_{ win}$(K$^2$)}\\
    \cmidrule(r){2-4}
    &$\rm |EoR-PL|$ & $\rm |EoR-SP|$& $\rm |SP-PL|$ \\
    \hline
    PCA $\rm N_{FG}: 3$ & $1.2898 \times 10^{-06}$ & $1.2912 \times 10^{-06}$ & $3.7581 \times 10^{-08}$ \\
    FastICA $\rm N_{FG}: 3$ & $1.2898 \times 10^{-06}$ & $1.2912 \times 10^{-06}$ & $3.7581 \times 10^{-08}$ \\
    PCA $\rm N_{FG}: 9$ & $8.9147 \times 10^{-09}$ & $8.9109 \times 10^{-09}$ & $6.5164 \times 10^{-10}$ \\
    FastICA $\rm N_{FG}: 9$ & $8.9147 \times 10^{-09}$ & $8.9109 \times 10^{-09}$ & $6.5164 \times 10^{-10}$ \\
    PCA $\rm N_{FG}: 15$ & $3.2709 \times 10^{-07}$ & $2.7145 \times 10^{-07}$ & $1.6559 \times 10^{-07}$ \\
    FastICA $\rm N_{FG}: 15$ & $3.2662 \times 10^{-07}$ & $2.7146 \times 10^{-07}$ & $1.6545 \times 10^{-07}$ \\
    \hline
    \end{tabular}
    \caption{Total power in the EoR window of the absolute difference between the 2D power spectrum of EoR signal and the residuals after subtracting the foregrounds using different component analysis methods.}
    \label{tab:CAs_ps2d}
\end{table}

\begin{figure}
\centering
\includegraphics[width=0.98\hsize]{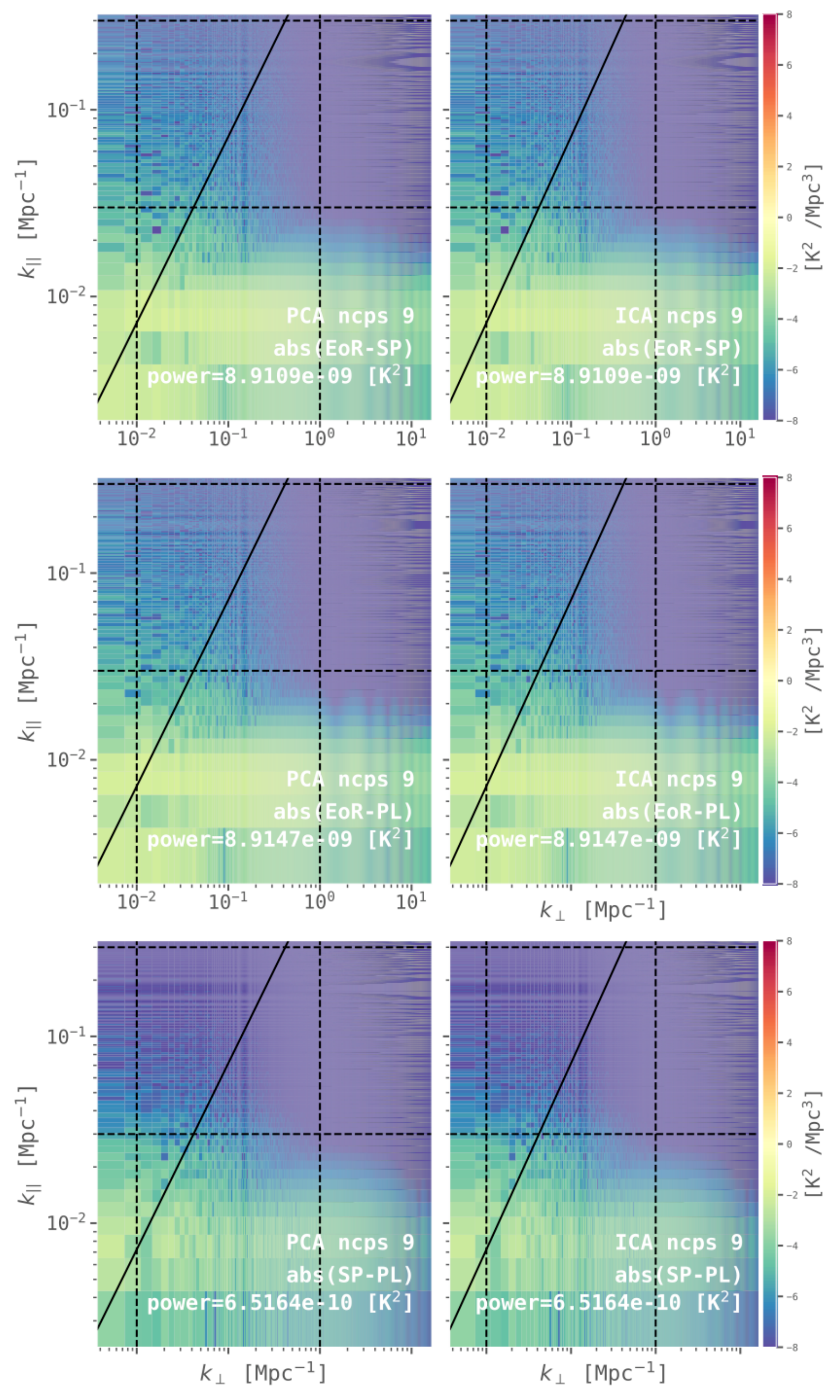}
\caption{The 2D power spectra of differences between the EoR signal and residuals obtained after removing the foreground with $\rm N_{FG}=9$. The results of PCA are displayed in the left panels, while the right panels showcase the results of FastICA. From top to bottom: $\rm PS_{EoR-SP}$, $\rm PS_{EoR-PL}$, and $\rm PS_{SP-PL}$. }
\label{fig:ps2d_CAs}
\end{figure}

We summarise $\rm P_{win}$ of the absolute difference between the EoR signal and the residuals in Table~\ref{tab:CAs_ps2d}.
It shows that PCA and FastICA produce similar performance with the same $\rm N_{FG}$, and when $\rm N_{FG}=9$, we obtain the $\rm P_{win}$ of residuals closest to those of the EoR signal (in comparison to $\rm N_{FG}=3$ and $\rm N_{FG}=15$).
In Figure~\ref{fig:ps2d_CAs}, we present $\rm PS_{|EoR-SP|}$, $\rm PS_{|EoR-PL|}$, and $\rm PS_{|SP-PL|}$ for the residuals obtained after subtracting the foreground with $\rm N_{FG}=9$.
It shows that the foreground residuals are still noticeable, but mainly confined to the foreground wedges.
Furthermore, the differences between residuals, i.e., $\rm P_{win, |SP-PL|}$, are much smaller than $\rm P_{win, |EoR-SP|}$ or $\rm P_{win, |EoR-PL|}$.
The $\rm P_{win}$ of the absolute difference between the residuals is 4.5 per cent of $\rm P_{win, |EoR-res|}$ and only approximately 0.01 per cent of $\rm P_{win, EoR}$.

\section{Conclusion and Discussion}
\label{sec:discussion}

In this work, we construct combined frequency spectra for sources from the GLEAM catalogue by cross-matching them with the RACS, NVSS, and SUMSS catalogues, and identify SP sources from these combined spectra.
We fit each source spectrum with the PL model (Eq.~\ref{eq:PL}), GC model (Eq.~\ref{eq:GC}), FFA model (Eq.~\ref{eq:FFA}) and CPL model (Eq.~\ref{eq:CPL}) .
SP sources are identified based on the presence of one of the following characteristics: a reliable spectral peak at a frequency greater than 130\,MHz or significant curvature within the GLEAM band. The final sample comprises 4,423 SP source candidates, representing approximately 3.2 per cent of the GLEAM radio source population.

We assess the impact of SP sources on EoR detection through mock observations. We generate the extragalactic sources sample and the EoR signal (from the 'Evolution Of 21 cm Structure' project) in the frequency range of 50 to 200\,MHz with a frequency resolution of 0.5\,MHz. We assume a perfect Gaussian synthesized beam with an angular resolution corresponding to an 80-km baseline.

For comparison, we consider two cases for the simulated extragalactic sources: (1) 96.8 per cent $S^3$ \citet{Wilman2008} power-low sources + 3.2 per cent SP sources (3.2\%-SP), and (2) 100 per cent power-low sources (3.2\%-PL) for comparison. We then compare the results of these two mocks using the same foreground removal methods to see the influence of 3.2 per cent of SP sources on foreground removal.

For the polynomial fitting method, the fit depends on both fitting bandwidth and polynomial order. Specifically, we use a $7^{\rm th}$-order polynomial for the spectrum in the 50--65\,MHz range and a $5^{\rm th}$-order polynomial for each spectrum within the 65--200\,MHz range with a bandwidth of 15\,MHz. Additionally, we employ a $9^{\rm th}$-order polynomial for the spectrum in the 50--80\,MHz range and a $7^{\rm th}$-order polynomial for each spectrum within the 80--200\,MHz range with a bandwidth of 30\,MHz.
After removing the foreground, a slight discrepancy can be observed at the level of approximately $10^{-5}$ K between the residual maps, i.e., the recovered EoR signal, of the two mock observations (3.2\%-SP and 3.2\%-PL). 
We quantify these differences in the angular power spectrum, and our findings indicate that the contribution from SP sources is less than 0.033 per cent of the EoR signal and less than 0.5 per cent of the difference between the EoR signal and the residuals of the 3.2\%-PL mock.


Furthermore, we compared the total power in the EoR window ($\rm P_{\text{win}}$) of the residuals. 
The differences between the EoR signal and the foreground-removed residuals ($\rm P_{\text{win, |EoR-SP|}}$ and $\rm P_{\text{win, |EoR-PL|}}$) are approximately 11 per cent (30\,MHz bandwidth) and 24 per cent (15\,MHz bandwidth) of $\rm P_{\text{win, EoR}}$.
Meanwhile, we observed extremely slight discrepancies between the residuals of the two simulations ($\rm P_{\text{win, |SP-PL|}}$) of about 0.06 per cent (30\,MHz bandwidth) and 0.04 per cent (15\,MHz bandwidth) of the EoR signal ($\rm P_{\text{win, EoR}}$). 
In other words, the $\rm P_{\text{win}}$ of residuals from extragalactic sources is approximately 2 to 3 orders of magnitude higher than the additional power in the EoR window introduced by SP sources.

In addition, we also compare the difference between the residuals after foreground removal with PCA and FastICA. Our fitting result shows that, with the same $\rm N_{FG}$, PCA and FastICA have similar performance in foreground removal, the $\rm P_{ win}$ of residuals are the same, and the residuals of foregrounds are mainly confined to foreground wedge. With $\rm N_{FG} = 9$, the $\rm P_{ win, |SP-PL|}$ only 0.01 per cent of $\rm P_{ win, EoR}$. 
In summary, all these results indicate that the impact of SP sources (about 3.2 per cent) is negligible with proper foreground removal setting if only SP sources and power-law spectrum sources are considered in the foreground.


However, these results are based on the SP sources we identified, which may be incomplete because of the lack of observations with frequencies less than 72\,MHz and faint sources. Future radio surveys at low frequencies such as SKA-low will provide more information about SP sources. Furthermore, our mock observations are simplified, we do not consider complex beam patterns, instrumental effects of telescopes, or other foreground effects, such as galactic free-free emission, and ionospheric influence in our work. If these effects are included in the mock observations, the frequency spectra would introduce more fluctuations, which would obscure the unsmoothness arising from SP sources. We defer more precise investigations into the impact of SP sources to future research. 

\section*{Acknowledgements}
We would like to extend our sincere appreciation to the anonymous reviewers for their valuable feedback and insightful comments, which have significantly improved the quality of this paper.
We would also like to express our heartfelt gratitude to all those who have contributed to this research. We acknowledge the support from the National SKA Program of China (no. 2020SKA0110100, 2020SKA0110200). QZ and HYS acknowledge the support from the National Science Foundation of China (11973069, 11973070). HYS acknowledges the support from Key Research Program of Frontier Sciences, CAS, Grant No. ZDBS-LY-7013.

\section*{Data Availability}
All the data and software packages used to produce the analysis in this article are publicly available (see Section~\ref{sec:data} and ~\ref{sect:results})



\bibliographystyle{mnras}
\bibliography{reference} 

\bsp	
\label{lastpage}
\end{document}